\documentclass[aps,prl,superscriptaddress,12pt]{revtex4-1}
\usepackage{graphicx}
\usepackage{times}
\usepackage{amsmath,amssymb}
\usepackage{bm}

\begin{document}

\title{Giant anomalous Nernst effect and quantum-critical scaling in a ferromagnetic semimetal}

\author{Akito Sakai}
\affiliation{Institute for Solid State Physics, University of Tokyo, Kashiwa, Chiba 277-8581, Japan}
\affiliation{CREST, Japan Science and Technology Agency (JST), 4-1-8 Honcho Kawaguchi, Saitama 332-0012, Japan}
\author{Yo Pierre Mizuta}
\affiliation{Faculty of Mathematics and Physics, Kanazawa University, Kanazawa, Ishikawa 920-1192, Japan}
\affiliation{Center for Emergent Matter Science (CEMS), RIKEN, Hirosawa, Wako, Saitama 351-0198, Japan}
\author{Agustinus Agung Nugroho}
\affiliation{CREST, Japan Science and Technology Agency (JST), 4-1-8 Honcho Kawaguchi, Saitama 332-0012, Japan}
\affiliation{Faculty of Mathematics and Natural Sciences, Bandung Institute of Technology, Bandung, Jawa Barat 40132, Indonesia}
\author{Rombang Sihombing}
\affiliation{Faculty of Mathematics and Natural Sciences, Bandung Institute of Technology, Bandung, Jawa Barat 40132, Indonesia}
\author{Takashi Koretsune}
\affiliation{CREST, Japan Science and Technology Agency (JST), 4-1-8 Honcho Kawaguchi, Saitama 332-0012, Japan}
\affiliation{Center for Emergent Matter Science (CEMS), RIKEN, Hirosawa, Wako, Saitama 351-0198, Japan}
\author{Michi-To Suzuki}
\affiliation{CREST, Japan Science and Technology Agency (JST), 4-1-8 Honcho Kawaguchi, Saitama 332-0012, Japan}
\affiliation{Center for Emergent Matter Science (CEMS), RIKEN, Hirosawa, Wako, Saitama 351-0198, Japan}
\author{Nayuta Takemori}
\affiliation{Center for Emergent Matter Science (CEMS), RIKEN, Hirosawa, Wako, Saitama 351-0198, Japan}
\author{Rieko Ishii}
\affiliation{Institute for Solid State Physics, University of Tokyo, Kashiwa, Chiba 277-8581, Japan}
\author{Daisuke Nishio-Hamane}
\affiliation{Institute for Solid State Physics, University of Tokyo, Kashiwa, Chiba 277-8581, Japan}
\author{Ryotaro Arita}
\affiliation{CREST, Japan Science and Technology Agency (JST), 4-1-8 Honcho Kawaguchi, Saitama 332-0012, Japan}
\affiliation{Center for Emergent Matter Science (CEMS), RIKEN, Hirosawa, Wako, Saitama 351-0198, Japan}
\author{Pallab Goswami}
\affiliation{Condensed Matter Theory Center and Joint Quantum Institute, Department of Physics, University of Maryland, College Park, Maryland 20742-4111, USA}
\affiliation{Department of Physics and Astronomy, Northwestern University, 2145 Sheridan Road, Evanston, IL 60208, USA}
\author{Satoru Nakatsuji}
\affiliation{Institute for Solid State Physics, University of Tokyo, Kashiwa, Chiba 277-8581, Japan}
\affiliation{CREST, Japan Science and Technology Agency (JST), 4-1-8 Honcho Kawaguchi, Saitama 332-0012, Japan}

\begin{abstract}

{\bf

In metallic ferromagnets, the Berry curvature of underlying quasiparticles can cause an electric voltage perpendicular to both magnetization and an applied temperature gradient, a phenomenon called the anomalous Nernst effect (ANE)\cite{Nagaosa2010,Niu2010}. 
Here, we report the observation of a giant ANE in the full-Heusler ferromagnet Co$_2$MnGa, reaching $S_{yx}\sim -6$ $\mu$V/K at room $T$, one order of magnitude larger than the maximum value reported for a magnetic conductor\cite{Hasegawa2015}. With increasing temperature, the transverse thermoelectric conductivity or Peltier coefficient $\alpha_{yx}$ shows a crossover between $T$-linear and $-T \log(T)$ behaviors, indicating the violation of Mott formula at high temperatures.
Our numerical and analytical calculations indicate that the proximity to a quantum Lifshitz transition between type-I and type-II magnetic Weyl fermions \cite{Wan2011,Burkov2011,type2} is responsible for the observed crossover properties and an enhanced $\alpha_{yx}$. The $T$ dependence of $\alpha_{yx}$ in experiments and numerical calculations can be understood in terms of a quantum critical scaling function predicted by the low energy effective theory over more than a decade of temperatures. Moreover, the observation of chiral anomaly or an unsaturated positive longitudinal magnetoconductance \cite{Nielsen1983,Son2013,Xiong2015} also provide evidence for the existence of Weyl fermions\cite{Wang2016,Kuebler2016} in Co$_2$MnGa. 
}

\end{abstract}

\maketitle

\newpage
Recent studies of novel phenomena arising from the coupling between spin and heat currents \cite{Uchida2008,Slachter2010,Huang2011} as well as new types of anomalous Hall effects in various magnets\cite{Nagaosa2010,Niu2010,Machida2010,Chen2014,Mn3Sn} have triggered renewed interest in the anomalous Nernst effect (ANE) as one of the topologically nontrivial phenomena and for its potential application to thermoelectric devices\cite{Nagaosa2010,Lee2004,Niu2010,Miyasato2007,Pu2008,Huang2011,Sakuraba2013,Hasegawa2015,IkhlasTomita2017,Kamran2017}.
ANE is known to generate an electric voltage \emph{perpendicular} to the applied temperature gradient $\vec{\nabla} T$ and magnetization $\vec{M}$, namely $\vec{E}_{\rm NE}=Q_{s}(\mu_0 \vec{M}\times\vec{\nabla} T)$, where $Q_s$ is the anomalous Nernst coefficient and $\mu_0$ is the vacuum permeability. This transverse geometry enables a lateral configuration of the thermoelectric modules to efficiently cover a heat source even with a curved surface\cite{Sakuraba2013}, having a much simpler structure than the modules using the conventional Seebeck effect \cite{Bell2008}. Since compared to the Seebeck effect, the ANE is a lesser studied phenomenon, there is an enormous scope for understanding the mechanism and controlling the size of ANE through new material synthesis. This should open a new avenue for identifying novel energy harvesting materials. 
 
On the other hand, the size of ANE in generic magnetic materials is too small for practical applications, and it is essential to overcome this hurdle. Promisingly, the recent theoretical and experimental investigations have indicated that the intense Berry curvature of Weyl points residing in the vicinity of the Fermi energy $E_{\rm F}$ can potentially enhance the intrinsic ANE\cite{Niu2010,IkhlasTomita2017,Kamran2017,Kuroda2017,Goswami2016}. However, there is still no clear analytical framework and a guiding principle for estimating and systematically increasing the size of the ANE for magnetic Weyl fermions by a few orders of magnitude. Therefore, experimental and theoretical studies of thermoelectric properties of Weyl magnets are critically important for both basic science and technological applications.

Recent first-principles calculations have showed that Co$_2$$TX$ ($T$ = transition metal, $X$=Si, Ge, Sn, Al, Ga) are potential magnetic Weyl metals, where multiple Weyl points exist in the momentum space near $E_{\rm F}$~\cite{Wang2016,Kuebler2016}. In particular, the first-principles calculations were performed to explain the experimentally observed giant AHE in the ferromagnet Co$_2$MnAl~\cite{Chen2004}. However, the large AHE does not guarantee a large ANE, because the ANE at low $T$s is given by the Berry curvature at $E_{\rm F}$, while the AHE is determined by the sum of the Berry curvature for all the occupied states~\cite{Niu2010,IkhlasTomita2017,Kamran2017}. 
Hence for a comprehensive understanding of the ANE of Weyl fermions and its correlation with AHE over a few decades of $T$s, we select the full Heusler ferromagnet Co$_2$MnGa, which has a Curie temperature $T_{\rm C}\sim 694$ K (Fig. 1a)\cite{Webster1971}. 

First, we describe our main result, namely the observation of a giant ANE in Co$_2$MnGa at room temperature. Figure 2a shows the magnetic field dependence of the Nernst signal $-S_{yx}$ for $\vec{B}\parallel$ [100], [110] and [111] and the heat current $\vec{Q}$ along [001] or [10$\bar{1}$]. Clearly, $-S_{yx}$ increases with elevating $T$ and reaches a record high value of $|S_{yx}|\sim 6$ $\mu$V/K at room temperature and it even approaches $\sim 8$ $\mu$V/K at 400 K (Fig. 2b), which is more than one order of magnitude larger than the typical values known for the ANE\cite{Hasegawa2015,IkhlasTomita2017,Kamran2017}. The observed value of $-S_{yx}$ is large in comparison with the Seebeck coefficient $S_{xx}$ (Fig. S2). For example, $|S_{yx}/S_{xx}|$ is 0.2, an unprecedented value for the Nernst angle $\theta_{\rm N} \approx$ tan$\theta_{\rm N} = S_{yx}/S_{xx}$ (Fig. 2a, right axis). In addition, we found there is almost no anisotropy in $S_{yx}$ within an error-bar (Supplementary Information).

Similar to the ANE, the Hall resistivity is found to be very large, reaching $\sim 15$ $\mu\Omega$cm at room temperature and its maximum $\sim 16$ $\mu\Omega$cm around 320 K (Figs. 2c and 2d). The Hall angle $ \theta_{\rm H} \approx$ tan$\theta_{\rm H} = \rho_{yx}/\rho_{xx}$ is also large and exceeds 0.1 at room temperature. Figures 2c and 2e show the field dependence of the Hall resistivity $\rho_{yx}$ and the magnetization $M$. Both the Hall and Nernst effects show nearly the same $B$ dependence as the magnetization curve, indicating that the anomalous contribution ($\propto M$) to the Hall and Nernst effects is dominant and the normal contribution ($\propto B$) is negligibly small at $T = 300$ K. The saturated magnetization, which is $M_{\rm s}\sim 3.8\mu_{\rm B}$ at $T = 300$ K, gradually grows on cooling and reaches $M_{\rm s}\sim 4\mu_{\rm B}$ at 5 K (Fig. 2f), consistent with the predicted value based on the Slater-Pauling rule. 
The anisotropy for $M$ is negligibly small at $T = 300$ K, which is fully consistent with the cubic structure. 

The observed $|\rho_{yx}| \sim 15$~$\mu\Omega$cm is one of the largest known for AHE. Likewise, the Hall conductivity is also exceptionally large. Figure 3a shows the $T$ dependence of the Hall conductivity, $\sigma_{yx}=-\rho_{yx}/(\rho_{xx}^2+\rho_{yx}^2)$, obtained at $B =2$ T. Here, $\rho_{xx}$ is the longitudinal resistivity, which is found to be isotropic as expected for a cubic system (Fig. S2). $-\sigma_{yx}$ monotonically increases on cooling and reaches $-\sigma_{yx}\sim 2000$ $\Omega^{-1}$cm$^{-1}$. This large value is of the same order of magnitude as the one known for the layered quantum Hall effect (QHE). Namely, the anomalous Hall conductivity can reach a value as large as $\sigma_{\rm H} =\frac{e^2}{h a} \sim 670~\Omega^{-1}$ cm$^{-1}$, a value expected for a 3D QHE with Chern number of unity, where $h$ is Planck constant and $a$ is the lattice constant\cite{Burkov2011}. 

We have also evaluated the $T$ dependence of the anomalous, transverse thermoelectric conductivity $\alpha_{yx}$ as shown in Fig. 3b (Supplementary Information). Up to $T \sim 25$ K, $-\alpha_{yx}$ increases almost linearly with $T$, displaying a maximum around $T\sim$140 K, followed by a gradual decrease. Notice that the $-\alpha_{yx}$ vs. $T$ curve closely resembles the functions $-T \log(T)$, and by plotting $-\alpha_{yx}/T$ against $\log(T)$ as shown in Fig. 3c, we find the crossover between two distinct scaling behaviors $\alpha_{yx} \sim T$ (at low $T$s) and $\alpha_{yx} \sim - T \log(T)$ (at high $T$s). The $T$-linear behavior is consistent with the Mott relation between $\alpha_{yx}$ at low $T$s ($k_B T \ll E_F$) and the energy derivative of $\sigma_{yx}$ at $T=0$, which predicts \begin{math}
        \alpha_{yx} \approx -\frac{\pi^2}{3}\frac{k^2_BT}{|e|}\frac{\partial \sigma_{yx}}{\partial E_{\rm F}}.
\end{math} However, the observed $\alpha_{yx} \sim -T \log(T)$ behavior over a decade of $T$s between $\sim 30$ K and 400 K constitutes a clear violation of Mott relation. In the inset of Fig. 3b, we have also presented the numerical calculations for $-\alpha_{yx}$ as a function of $T$. While the maximum value of $-\alpha_{yx}$ and the overall functional dependence on $T$ are in considerable agreement, the temperature scales where $-\alpha_{yx}$ attains its maximum value have an order of magnitude difference. Now we show how these results can be understood in terms of Weyl fermions.

The anomalous Hall conductivity at $T=0$ is given by
$
        \sigma_{yx}(E_{\rm F})=-\frac{e^2}{\hbar}\sum_{n,\bm{k}} \Omega_{n,z}(\bm{k}) \times \theta(E_{\rm F}-\epsilon_{n,\bm{k}}),
$
involving the summation over all occupied states, where $\Omega_{n,z}(\bm{k})$ is the Berry curvature along the $\hat{z}$ direction, $n$ is the band index, and $\theta(x)$ is the unit step function.
Consequently,
\begin{math}
        \frac{\partial \sigma_{yx}}{\partial E_F}=-\frac{e^2}{\hbar}\sum_{n,\bm{k}} \Omega_{n,z}(\bm{k}) \: \delta(E_{\rm F}-\epsilon_{n,\bm{k}}),
\end{math}
where $\delta(x)$ is the Dirac delta function, and only partially occupied bands can contribute to $\alpha_{yx}$.  Therefore, to obtain a large $\alpha_{yx}/T$, it is essential to concomitantly enhance the density of states (DOS) and the Berry curvature around the Fermi pockets. 

Thus, we focus on the largest Fermi surface (Fig. 1b, Supplementary Information, Fig. S4) and find that the pertinent bands in the numerical calculations produce Weyl points around $E_0 \approx +20$ meV above $E_{\rm F}$\cite{Kuebler2016} (Fig. 1c). Indeed, this Fermi surface has a large Berry curvature due to its proximity to the Weyl points (as shown in Fig. 1d) and also a large DOS (Fig. S5) due to the flatness of the dispersion. These Weyl points are located on the zone boundary along U-Z-U line (Fig. 1d) at
 $\pm \bm{k}_0= \pm \frac{2\pi}{a} \times 0.15~[110]$ and along the nodal direction they can be modeled with the low energy Hamiltonians, 
\begin{math}
H_j \approx \mathrm{sgn}(j)[\hbar v_2 (\bm{k}-\mathrm{sgn}(j)\bm{k}_0)\cdot\hat{\bm{k}}_0 + \hbar v_1 (\bm{k}-\mathrm{sgn}(j)\bm{k}_0)\cdot\hat{\bm{k}}_0 \sigma_z], 
\end{math}
where $j = \pm1$ denote the chirality of the right and left handed Weyl points, $v_1$ and $v_2$ are two velocity parameters with the tilt parameter $v_2/v_1 = 0.99$, and $v_1 \approx 10^5$m/s. 
The strength of the tilt parameter $v_2/v_1$ is extremely close to the critical value $v_2/v_1=1$, which describes a Lifshitz quantum critical point (LQCP), separating the type-I Weyl fermions ($v_2/v_1 < 1$, with a vanishing DOS at the Weyl points, causing only an electron or a hole pocket) from the type-II Weyl fermions ($v_2/v_1 > 1$, with a finite DOS at the Weyl points, where the electron and hole pockets touch). At the LQCP, the energy derivative of the Hall conductivity displays the following singular behavior 
\begin{equation}
\frac{\partial \sigma_{yx}}{\partial E} \approx -\frac{e^2}{4\pi^2 \hbar^2 v_1} \; \log \left(\frac{|E-E_0|}{C(k_0)\hbar v_1 k_0}\right), \: C(k_0)=\frac{8 \sin(k_0a/2)}{k_0a} \exp[-4 \tan(k_0 a/4)]. \label{scaling1}
\end{equation}
Away from the LQCP (inside type-I or type-II phases) this log divergence gets cut off by the distance from the LQCP defined as $\delta= 1-v_2/v_1$, leading to $\frac{\partial \sigma_{yx}}{\partial E} \propto  \log |\delta|$~(Supplementary Information). Since $\delta \hbar v_1 k_0 < |E_F-E_0| < \hbar v_1 k_0$, these Weyl fermions belong to the quantum critical regime of LQCP. The numerically calculated $\frac{\partial \sigma_{yx}}{\partial E}$ is shown in Fig. 3e. By analyzing the sharp peak around $E_0 \approx +20$ meV, we have clearly identified the log divergent behavior in Fig. 3f.

Intriguingly, the low energy theory suggests that the effects of LQCP over a wide range of temperatures can be captured in terms of a scaling function 
\begin{equation}\alpha_{yx} (T, \mu) = \frac{k^2_B |e| T_0}{12 \hbar^2 v_1}  \; G \left(\frac{T}{T_0}, \frac{\mu-E_{0}}{k_BT_0} \right),
\label{scaling2} \end{equation} where $T_0 \approx \exp[1] \times T_m$, and $T_m$ is the temperature where $\alpha$ attains its maximum value $\alpha^{\rm max}_{yx}$. By setting $\mu=E_0$, we have obtained the scaling function at the LQCP as
\begin{eqnarray}
\alpha_{yx}(T,0)&=& \alpha^{\rm max}_{yx}(k_0) \; \exp[1] \; \frac{T}{T_0} \log \left(T/T_0 \right), \; k_BT_0 \approx C(k_0) \hbar v_1 k_0,  \nonumber \\ \; \alpha^{\rm max}_{yx}(k_0)&=&\frac{k^2_B e T_0}{12 \hbar^2 v_1 \exp[1]} \approx \frac{50}{a} \sin \left(\frac{k_0a}{2} \right) \exp \left[-4 \tan \left(\frac{k_0a}{2}\right)\right] {\rm AK^{-1}m^{-1},} \label{scaling3}
\end{eqnarray}
where the lattice constant is measured in angstroms. Notice that the maximum value $\alpha^{\rm max}_{yx} (k_0)$ does not depend on the velocity $v_1$ of the Weyl fermions. In contrast, the slope $\alpha_{yx}/T \sim -(k^2_B |e|)/(12 \hbar^2 v_1)$ explicitly depends on $v_1$. We have found that $\alpha^{\rm max}_{yx} (k_0)$ is maximized for $k_0 \sim 0.14 \times (2\pi/a)$ and the numerically determined location of Weyl points is indeed very close to this value. After substituting $a=5.8 \times 10^{-10}$ m and $k_0=0.15 \times (2\pi)/a$ we find $\alpha^{\rm max}_{yx} \approx 1.5$ AK$^{-1}$m$^{-1}$, which is about $3$ times smaller than the experimentally and numerically obtained values. We note that the estimation for $T_0$ can be slightly modified by microscopic details and even after ignoring the factor $C(k_0)$ we would obtain $\alpha^{\rm max}_{yx} \approx 1$ AK$^{-1}$m$^{-1}$. Such $O(1)$ uncertainty regarding the value of $T_0$ and the presence of a few additional pairs of Weyl points in the vicinity of the Fermi level can lead to the $O(1)$ difference between this estimate and the experimentally and numerically determined values of $\alpha^{\rm max}_{yx}$.

We have also computed $\alpha_{yx}(T,\mu)$ for $\mu \neq E_0$, in order to demonstrate its crossover form $T \log(T)$ behavior to $\alpha_{yx} \approx -\alpha^{\rm max}_{yx} \; \frac{T}{T_0} \; \log \left (\frac{|\mu-E_0|}{k_BT_0} \right),$ when $k_B T \leq |\mu-E_0|$ (the solid black line in Fig. 3c). In the real material, the scale $T_0$ (proportional to $v_1$) is an order of magnitude smaller than the numerical one, which we can attribute to the correlation and disorder driven suppression of the bandwidth or the velocity of Weyl fermions. This order of magnitude suppression is consistent with experimentally found enhancement of the Sommerfeld coefficient for the specific heat by a factor of $7$ (Supplementary Information). Therefore, the experimental and numerical values for $\alpha_{yx}/T$ differ by an order of magnitude. 
Finally in Fig. 3c we have compared the dimensionless scaling function $G$ obtained from the experimental and numerical data with the predictions of low energy theory. After selecting $T_0=550$ K and $T_0=6000$ K respectively for the experimental and numerical results, $G$ function matches with the low energy predictions for more than a decade of $T$s. The crossover to Mott regime occurs when $(\mu-E_0)/(k_B T_0) \approx -0.05$. Thus, our analysis indicates that the proximity of the Weyl fermions to a quantum Lifshitz transition is responsible for the large values of $\alpha^{\rm max}_{yx}$, the slope $\alpha_{yx}/T$ at low $T$s, as well as the logarithmic violation of Mott relation at high $T$s.

To provide further evidence for the existence of Weyl fermions in the vicinity of $E_{\rm F}$, we have performed angle dependent magnetoresistance measurements to reveal the chiral anomaly. In Weyl metals, the number imbalance between the Weyl nodes with opposite chirality is expected to cause negative longitudinal magnetoresistance when the electric current $\vec{I}$ and magnetic field $\vec{B}$ are parallel ($\vec{I} \parallel \vec{B}$), while the transverse magnetoresistance remains positive ($\vec{I} \perp \vec{B}$)\cite{Nielsen1983,Son2013,Xiong2015}. However, such behavior can be masked in a ferromagnetic Weyl metal for weak magnetic fields, since field induced suppression of magnetic fluctuations (decrease in scattering rate) can cause positive magnetoconductance for an arbitrary angle between $\vec{I}$ and $\vec{B}$. Therefore, it is imperative to perform the measurements at sufficiently low $T$s and in the presence of strong enough magnetic fields to minimize the effects of magnetic fluctuations. Figure 4 and Fig. S6 show the $B$ dependence of the longitudinal and transverse magnetoconductivity $\sigma_{xx}(B)$ at $T$ = 5 K and in $B \leq 16$ T with different current directions (Supplementary Information). Being fully consistent with this idea, we find clear signature of chiral anomaly (positive and negative magnetoconductivities respectively for $\vec{I} \parallel \vec{B}$ and $\vec{I} \perp \vec{B}$ (Fig. 4a), and their $\cos^2(\theta)$ dependence on the angle $\theta$ between $\vec{I}$ and $\vec{B}$ (Fig. 4b)) in high magnetic fields (e.g. above $|\vec{B}| \sim 6 \; \mathrm{T}$ for $\vec{I} \parallel$ [100] as shown in Fig. 4a)(Supplementary Information). The chiral anomaly has been observed down to $T=0.1$ K, providing evidence that the Weyl excitations remain gapless at very low $T$s even in the presence of strong electronic interactions.

In summary, our comprehensive study indicates that the giant Nernst effect in Co$_2$MnGa originates from the large intrinsic transverse thermoelectric conductivity $\alpha_{yx}$ of Weyl points occurring near $E_{\rm F}$. In fact, $|\alpha_{yx}|\sim4$ AK$^{-1}$m$^{-1}$ for Co$_2$MnGa is very large compared to the other typical ferromagnets $|\alpha_{yx}|\sim 0.01$-1 AK$^{-1}$m$^{-1}$ as shown in Fig. 4c (Supplementary Information). Our combined experimental, numerical and effective theory based analysis provides a guideline for further increasing $\alpha_{yx}$ in Weyl magnets, by decreasing the lattice constant. The observed large ANE with small anisotropy for Co$_2$MnGa indicates that Weyl magnets are potentially useful to create efficient thermopile devices for the thermoelectric power generation as well as to study the quantum critical effects of Weyl fermions in spintronics (Supplementary Information). 

\section*{Methods}
\section*{Co$_2$MnGa single crystal and experimental methods}
Single crystals of Co$_2$MnGa were prepared by the Czochralski method after making polycrystalline samples by arc-melting Co, Mn, and Ga with an appropriate ratio.
Co$_2$MnGa is known to be highly resistive to oxidation and we found that the sample is stable in air\cite{Ludbrook2017}. As-grown single crystals were used for all the measurements except powder X-ray diffraction as described below. Our analyses using both the inductively coupled plasma (ICP) spectroscopy and energy dispersive X-ray analysis (EDX) indicate that our single crystals are stoichiometric within a few \% resolution. 

The samples were oriented by the Laue backscattering method, and then cut into a bar-shape by spark erosion. All the surfaces were polished to get flat and mirror-like surfaces. 
Three samples were prepared by the same procedure to perform the measurements in the different geometries, $\#100$ for $\vec{B}\parallel$ [100], $\#110$ for $\vec{B}\parallel$ [110] and $\#111$ for $\vec{B}\parallel$ [111]. All the three samples have nearly the same dimension with the typical size of $7.5 \times 2.0 \times 1.3$ mm$^3$. For both Seebeck and Nernst effect measurements, the distance between the thermometers $l_{\rm th}$ was set to be $\sim 4.0$ mm. Given the width of the temperature probes, the error-bar of the Seebeck and Nernst signals mainly comes from the uncertainties of the corresponding geometrical factors and is estimated to be $\sim 10$ \%.
All the transport measurements including the electric resistivity, Hall, Seebeck and Nernst effects as well as thermal conductivity were measured for each bar-shape sample using a commercial system (PPMS, Quantum Design). 
Magnetization was measured for a small piece of the single crystal by using a commercial SQUID magnetometer (MPMS, Quantum Design). The sample was reshaped into a cubic-like shape to reduce the shape anisotropy, before the magnetization measurements. 

\section*{Data availability}
The data that support the plots within this paper and other findings of this study are available from the corresponding author upon reasonable request.

\section*{Acknowledgements }
 This work was supported by CREST (JPMJCR15Q5) by Japan Science and Technology Agency, by Grants-in-Aid for Scientific Research (Grant Nos. 16H02209, 25707030), by Grants-in-Aid for Scientific Research on Innovative Areas “J-Physics” (Grant Nos. 15H05882 and 15H05883) and Program for Advancing Strategic International Networks to Accelerate the Circulation of Talented Researchers (Grant No. R2604) from the Japanese Society for the Promotion of Science. P. G. was supported by JQI-NSF-PFC and LPS-MPO-CMTC (at the University of Maryland) and the start-up funds from the Northwestern University. The use of the facilities of the Materials Design and Characterization Laboratory at the Institute for Solid State Physics is appreciated.
\section*{Contributions]}
S. N. conceived and planned the experimental project. A. N., R. S., S. N. worked on the single crystals growth and the preparation of samples. A. S., R. S. carried out the transport and low temperature measurements and analyzed the data. Y. M., T. K., M. S., N. T., R.A. performed the first-principles calculations. P. G. formulated the quantum critical theory and the scaling analysis of experimental and numerical results. S. N. made the scaling analysis. R. I. made the chemical analyses. D. H. took the electron diffraction image. S. N., A. S. and P. G. wrote the paper with inputs from Y. M., and R. A.. All authors discussed the results and commented on the manuscript.
\section*{Competing Interests}
The authors declare that they have no
competing financial interests.
\section*{Corresponding authors}
Correspondence and requests for materials should be addressed to S.N.
(email: satoru@issp.u-tokyo.ac.jp).


\newpage
\begin{figure}
\begin{center}
\includegraphics[keepaspectratio, scale=0.45]{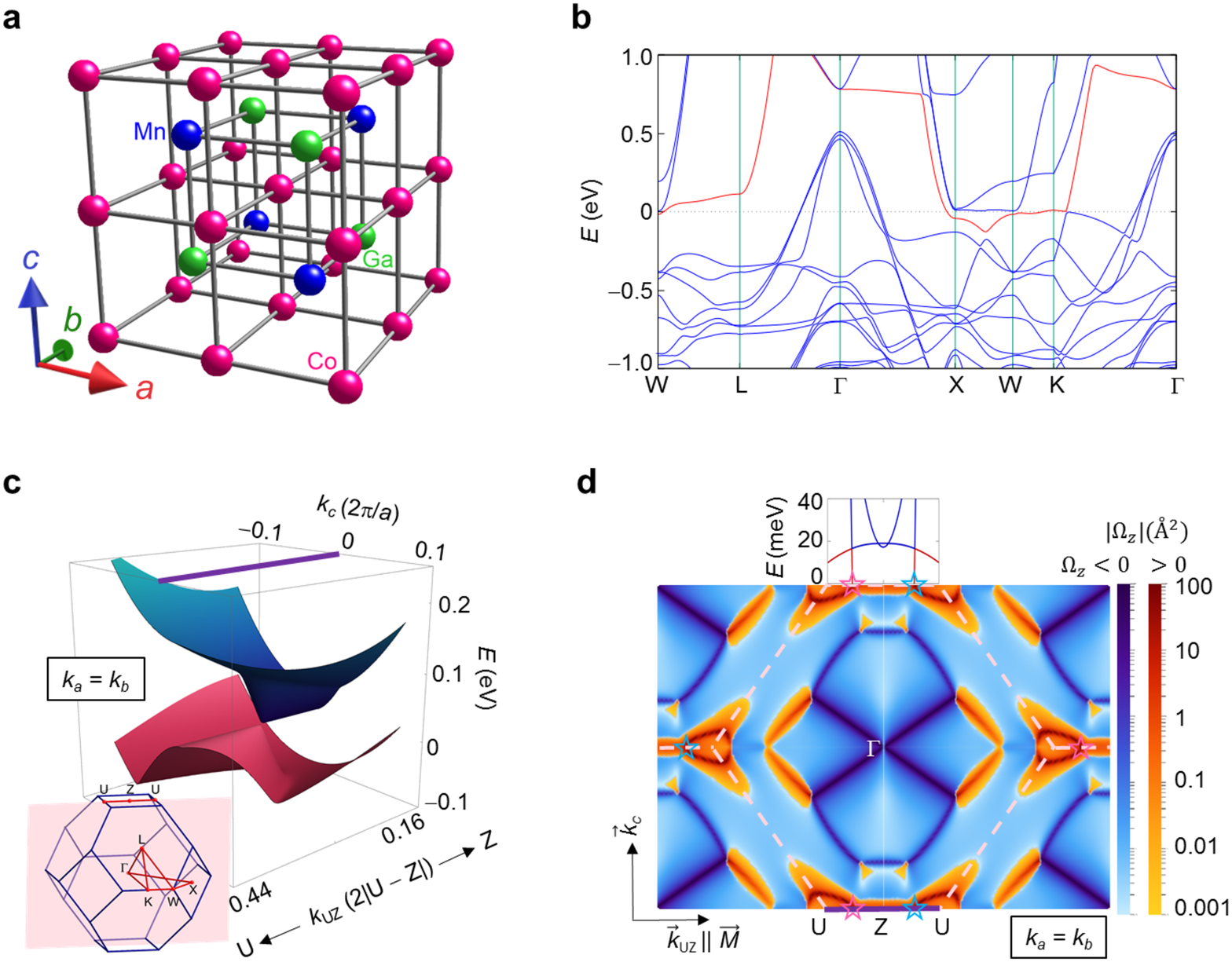}
\end{center}
\end{figure}
\normalsize{{\bf Figure 1 $|$ Crystal structure, theoretical band structure and Weyl points of Co$_2$MnGa.} {\bf a}, $L2_1$ ordered cubic full Heusler structure, which consists of four fcc sublattices, confirmed by the X-ray and electron diffractions (Supplementary Information). {\bf b}, Band structure of Co$_2$MnGa obtained from the first-principles calculations for the case of magnetization $M$ = 4.2 $\mu_{\rm B}$ along [110]. The band which forms the largest Fermi surface is colored in red. {\bf c}, Weyl points located along U-Z-U line in the $k_a = k_b$ plane spanned by the momentum $k_{\rm UZ}$ along U-Z and $k_c$. A higher energy (red) and a lower energy (blue) non-degenerate bands touches at the point with a linear dispersion. The tilt parapeter $v_2/v_1$ is very close to unity, indicating the proximity of the quantum Lifshitz transition. The inset indicates the first Brillouin zone and symmetric points of fcc lattice. The $k_a = k_b$ plane is shown by the pink plane. {\bf d}, The $z$ component of the Berrry curvature $\Omega_z$ in the $k_a = k_b$ plane (bottom panel) and band structure along U-Z-U at $E\sim 20$ meV (top panel). Here, $\hat{z}$ direction is taken to be the quantization axis ($\hat{z}\parallel \vec{M}$). The deep-pink and skyblue stars in the bottom panel represent the positive and negative Weyl points, respectively. The red-colored band dispersion curve in the top panel is identical to those shown by the red color in Figs. 1b and 1c.
}

\newpage
\begin{figure}
\begin{center}
\includegraphics[keepaspectratio, scale=0.8]{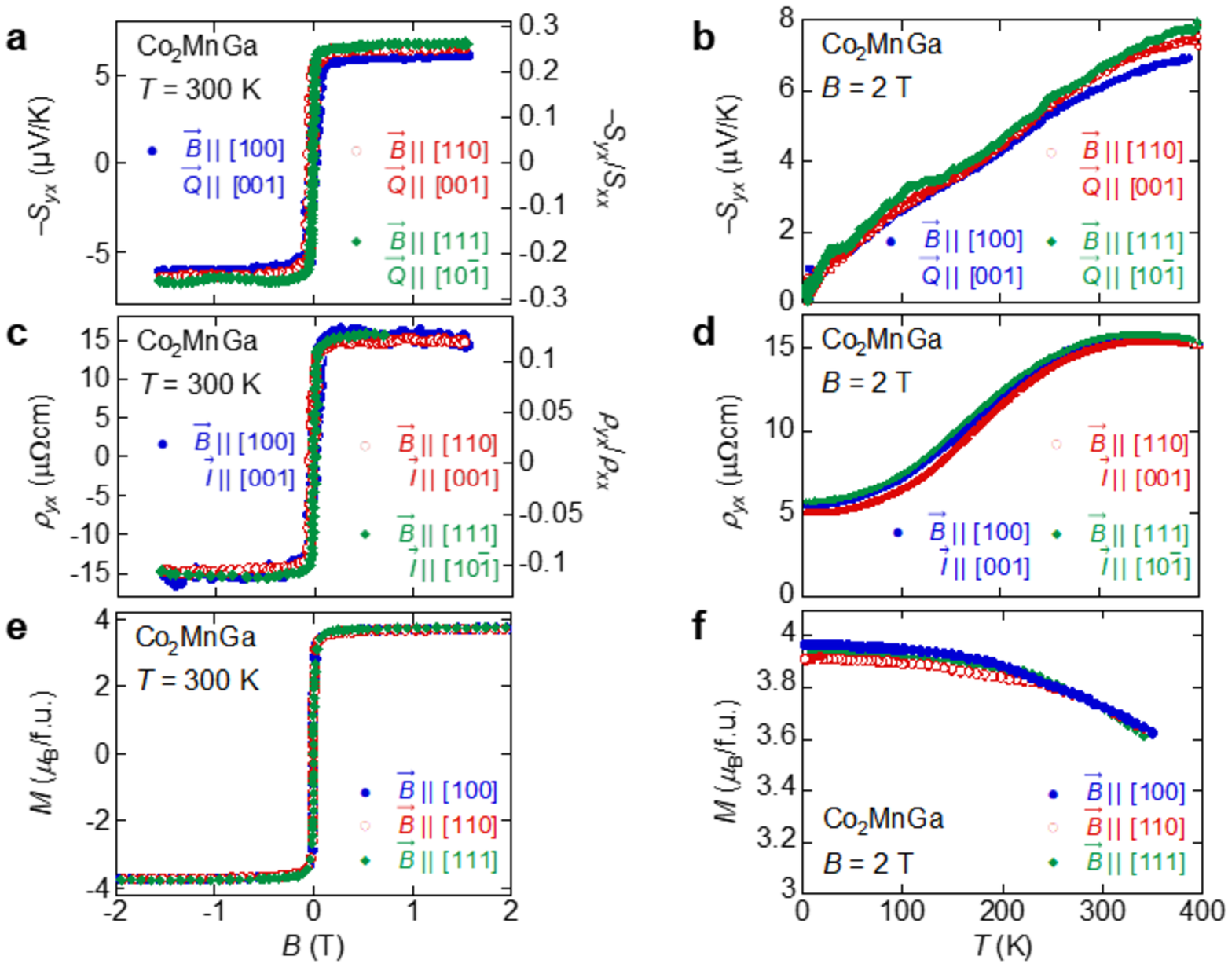}
\end{center}
\end{figure}
\normalsize{{\bf Figure 2 $|$ Observation of the giant anomalous Nernst effect at room temperature in Co$_2$MnGa.} {\bf a}, {\bf b}, Nernst signal $S_{yx}$ as a function of magnetic field $\vec{B}$ ({\bf a}) and temperature $T$ ({\bf b}). {\bf c}, {\bf d}, Hall resistivity $\rho_{yx}$ as a function of $\vec{B}$ ({\bf c}) and $T$ ({\bf d}). {\bf e}, {\bf f}, Magnetization $M$ as a function of $\vec{B}$ ({\bf e}) and $T$({\bf f}). All the $\vec{B}$ and $T$ dependence data are taken at room $T$ and $|\vec{B}|=2$ T, respectively, in $\vec{B}\parallel$ [100] (solid circle), [110] (open circle) and [111] (solid diamond). The magnitudes of the Nernst angle $-S_{yx}/S_{xx}$ and the Hall angle $\rho_{yx}/\rho_{xx}$ are shown in the right axes of the panels {\bf a} and {\bf c}, respectively. The magnitude of the magnetic field along the horizontal-axis has been corrected for the demagnetization effect. }

~~

\begin{figure}
\begin{center}
\includegraphics[keepaspectratio, scale=0.6]{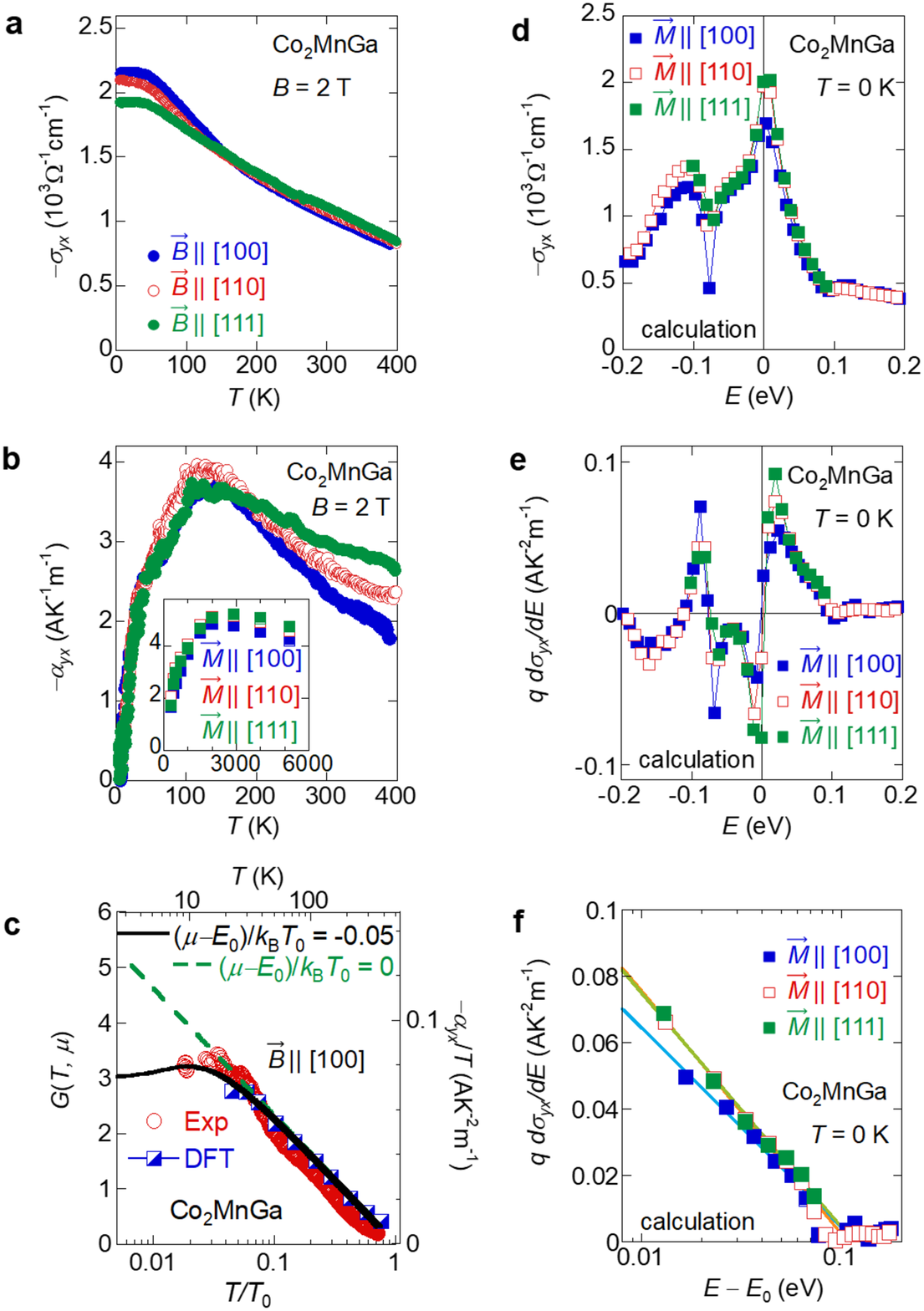}
\end{center}
\end{figure}
\newpage

\normalsize{{\bf Figure 3 $|$ Giant anomalous Hall and transverse thermoelectric conductivities and the crossover between the regimes following and violating the Mott relation.} {\bf a}, {\bf b}, Temperature dependence of the Hall conductivity $\sigma_{yx}$ ({\bf a}) and the transverse thermoelectric conductivity $\alpha_{yx}$ ({\bf b}) measured in a field of $|\vec{B}| = 2$ T along [100], [110], and [111]. Inset: Temperature dependence of $\alpha_{yx}$ obtained by DFT methods for the states having the magnetization $\vec{M}$ parallel to [100], [110] and [111]. {\bf c}, Dimensionless scaling function of Eq. (2) $G(T,\mu)$, (left vertical axis) vs. $T/T_0$ (lower horizontal axis) obtained for the Nernst measurement (circle, $T_0 =550$ K) in a field of $|\vec{B}| = 2$ T along [100] and for DFT calculations (square, $T_0 = 6000$ K) for the states having the magnetization $\vec{M}$ parallel to [100]. $G$ functions for experiment and DFT calculations match with the results (solid line) for the low energy model over a decade of $T$s. The dashed line is the quantum critical scaling function from Eq.~(3) when the chemical potential $\mu$ is tuned at the Weyl points, and the unbounded, logarithmic growth of critical $G$ function at low temperatures describes the critical enhancement of $\alpha_{yx}/T$ and breakdown of Mott relation. Above a crossover temperature determined by $\mu$, the $G$ function from experiments, DFT calculations, and low energy results with ($\mu-E_0)/k_{\rm B}T_0=-0.05$ (solid line) follow the quantum critical result. For experiment, $\alpha_{yx}/T$ (right vertical axis) vs. $T$ (upper horizontal axis) (Supplementary Information).  {\bf d}, {\bf e}, Anomalous Hall conductivity $\sigma_{yx}$ ({\bf d}) and the energy derivative of $\sigma_{yx}$ at zero temperature ({\bf e}) for the states having the magnetization $\vec{M}$ parallel to [100], [110], and [111] obtained by first-principles calculations (Supplementary Information), with $q = \frac{\pi^2}{3}\frac{k^2_B}{|e|}$. According to the Mott relation, at sufficiently low temperatures $\alpha_{yx}/T=-q \frac{\partial \sigma_{yx}}{\partial E_{\rm F}}$. {\bf f,} $-\ln |E-E_0|$ dependence of $\frac{\partial \sigma_{yx}}{\partial E_{\rm F}}$ shown in ({\bf e}) in the vicinity of the Weyl point (Fig. 1c, d) at $E_0 \approx + 0.02$ eV above the Fermi energy, in accordance with the predictions of low energy theory, described in Eq.~(1). 
 }

\newpage
\begin{figure}
\begin{center}
\includegraphics[keepaspectratio, scale=0.5]{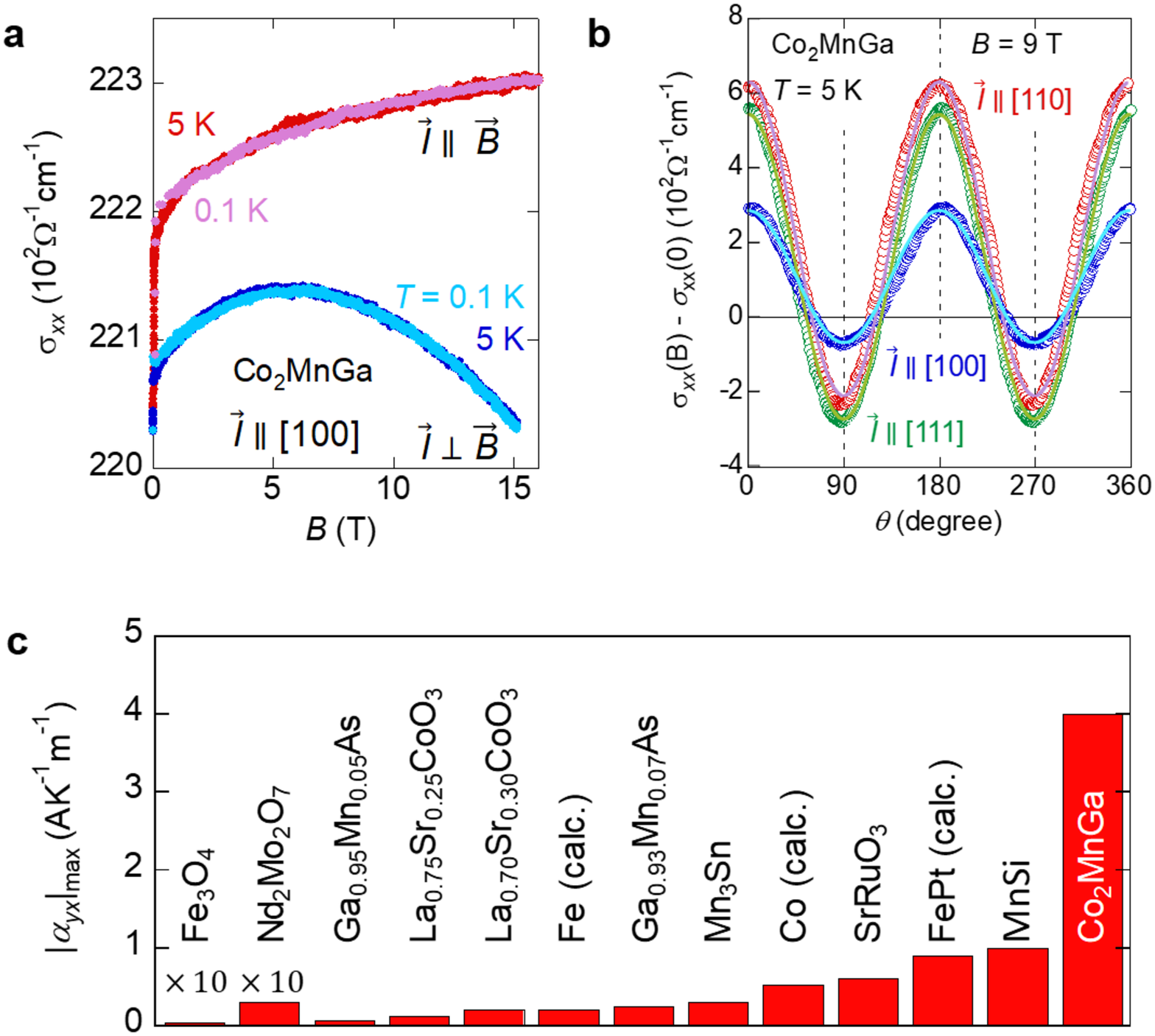}\label{MR}
\end{center}
\end{figure}
\normalsize{{\bf Figure 4 $|$ Evidence for the Weyl metal state in the ferromagnetic Co$_2$MnGa.} {\bf a}, Magnetic field dependence of the longitudinal electric conductivity $\sigma_{xx}$ at $T$ = 5 K and 0.1 K for $\vec{I} \parallel \vec{B}$ and $\vec{I} \perp \vec{B}$ for Co$_2$MnGa. {\bf b}, Angle $\theta$ dependence of the magnetoconductance for $\vec{I} \parallel$ [100], [110] and [111] at $|\vec{B}| = 9$ T for Co$_2$MnGa. $\theta$ is the angle between the magnetic field and the electric current direction (Supplementary Information). $\theta = 0$ and 90$^{\circ}$ correspond to the configurations for $\vec{I} \parallel \vec{B}$ and $\vec{I} \perp \vec{B}$ in  {\bf a}, respectively. The solid lines indicate the fits to $\cos^2(\theta)$. {\bf c}, Magnitude of the transverse thermoelectric conductivity $|\alpha_{yx}|$ for various ferromagnets and Weyl antiferromagnet Mn$_3$Sn (Supplementary Information).
}

\section*{Supplementary Information}

\section*{Crystal structure}
Co$_2$MnGa has the cubic ordered full Heusler type structure ($L2_1$) with the space group $Fm\overline{3}m$. It consists of four fcc sublattices with Co atoms at (1/4, 1/4, 1/4) and (3/4,3/4,3/4), Mn atom at (0, 0, 0) and Ga atom at (1/2, 1/2, 1/2). It is known that this structure is prone to have disorder; forming either $B2$ structure with $Pm\overline{3}m$ symmetry, where Mn and Ga atoms are randomly interchanged, or $A2$ structure with $Im\overline{3}m$ symmetry, where all atoms are randomly mixing. Our single crystals were confirmed to have $L2_1$ structure, as clearly represented by the existence of 111 reflections in various diffraction methods: powder X-ray diffraction (XRD) (Fig. S1a), selected area electron diffraction (Fig. S1b) and single crystal XRD. The XRD pattern is well reproduced by the Rietveld analysis as shown by the solid line in Fig. S1a, indicating the single phase of the $L2_1$ cubic Heusler Co$_2$MnGa with the lattice constant of $a=5.77(3) \AA$ (Table S1). To make the powder XRD measurements, we annealed the sample at 700 $^\circ$C for $\sim 3$ hours after making powder from single crystals to remove the strains and distortions introduced by the crushing procedure. The 2D- and 1D- characteristic X-ray maps demonstrate the homogeneity of the stoichiometry over the entire sample  (Fig. S1c and S1d).

\section*{Transport properties and estimation of the transverse thermoelectric conductivity $\alpha_{yx}$}
Figure S2 shows the $T$ dependence of ({\bf a}) the electric resistivity, $\rho$, ({\bf b}) the Seebeck effect, $S_{xx}$, and ({\bf c}) thermal conductivity, $\kappa$, at zero field, $B=0$, for each bar-shaped sample labeled by the field directions applied during the Nernst and Hall effect measurements. Electric and heat current flows along [001] for $\#$100 and $\#$110 and along [10$\overline{1}$] for  $\#$111, respectively. All the $\rho$ vs. $T$ curves nearly collapse on top, indicating the small sample dependence and weak anisotropy in $\rho$. Over one decade between 2 and 40 K, the longitudinal resistivity $\rho_{xx}$ shows $T^2$ dependence as shown in the inset of Fig. S2a. The solid lines represent the fit to the equation $\rho_{xx}=\rho_0+AT^2$, which yields $A\sim8\times10^{-4}$ ($\mu\Omega$cm/K$^2$).

For $S_{xx}$, only the sample $\#$111 shows different behavior from the other two, indicating the heat current direction dependence of the Seebeck effect. Resistivity monotonically decreases with decreasing $T$ without any minimum, which excludes the possibility of the weak localization effect as discussed in the following section. The small peak in $S_{xx}$ at $\sim 90$ K can be attributed to the phonon drag since this temperature is close to the typical peak $T$ expected for the phonon drag\cite{Ziman1960}, i.e. $\sim T_{\rm D}/5$ using the Debye temperature $T_{\rm D}= 360 \pm 10$ K estimated in the following section of Supplementary Information.

Similar to the longitudinal resistivity $\rho$, the thermal conductivity $\kappa$ also shows small sample dependence and weak anisotropy. $\kappa(T)$ forms a peak around 70 to 90 K most likely following the  temperature dependence of the phonon thermal conductivity $\kappa_{\rm ph}$ as often observed in crystalline materials. Namely, assuming the relation $\kappa_{\rm ph}=C_{\rm ph}vl_{\rm ph}/3$, where $C_{\rm ph}$, $v$ and $l_{\rm ph}$ are specific-heat, mean group velocity and mean free path, respectively, it is conceivable that with decreasing $T$ from high temperature, $\kappa_{\rm ph}$ first increases  because $l_{\rm ph}$ increases with reducing the phonon-phonon scattering and then peaks with the salutation of $l_{\rm ph}$ and finally decreases roughly proportional to $C_{\rm ph}$ at low $T$s.

In general, electric current is generated by both electric field $\vec{E}$ and temperature gradient $\vec{\nabla} T$, namely
\begin{eqnarray*}
\vec{J}=\hat{\sigma}\vec{E}-\hat{\alpha}\vec{\nabla} T,
\end{eqnarray*}
where $\vec{J}$, $\hat{\sigma}$ and $\hat{\alpha}$ are the electric current 
density, the electric and thermoelectric (Peltier) conductivity tensors, 
respectively. Setting $\vec{B} \parallel \hat{z}$ and $\vec{\nabla} T \parallel \hat{x}$, and the open circuit condition $\vec{J}=0$, we obtain,
\begin{eqnarray*}
J_y&=&\sigma_{yx}S_{xx}+\sigma S_{yx}-\alpha_{yx}=0,
\end{eqnarray*}
where we assume that the Seebeck coefficient $\sigma=\sigma_{xx}=\sigma_{yy}$ for the cubic symmetry. Using this equation and values from the experiment, we calculate the $T$ dependence of $-\alpha_{yx}$ (Fig. 3b).

\section*{Specific heat}
Figure S3 shows the temperature dependence of the specific heat divided by $T$, $C/T$ obtained for Co$_2$MnGa at zero field. 
To estimate the Sommerfeld coefficient $\gamma$, $C/T$ vs. $T^2$ below $T\sim 20$ K is plotted in the inset of Fig. S3.
The broken line is the fit by $C/T=\gamma + \beta T^2$, where the first and second terms respectively indicate the electronic and low $T$ limit of Debye-type phonon contributions. Using the relation $\beta=(12\pi^4Nk_{\rm B}r)/(5T_{\rm D}^3)$, where $N$, $k_{\rm B}$, $r$ and $T_{\rm D}$ are the Avogadro's number, Boltzmann constant, number of formula unit per unit-cell and Debye temperature, respectively, we obtain $\gamma = 12.2$ mJ/(moleK$^2$) and $T_{\rm D}=366$ K, consistent with the previous report\cite{Umetsu2010}. Experimentally obtained $\gamma = 12.2$ mJ/(moleK$^2$) is found to be $\sim 7$ times larger than the value estimated from the first-principles calculation $\gamma = 2$ mJ/(moleK$^2$). This enhancement should come from the band renormalization due to the strong electron correlation among $3d$ conduction electrons, which is not taken into account in the first-principles calculation. Namely, this indicates the $\sim 6$ times shrinkage of the band width in Co$_2$MnGa than the one obtained from the band calculation. Note that Kadowaki-Woods ratio $A/\gamma^2$, where $A$ is the $T^2$ coefficient of the longitudinal resistivity as discussed in the previous section, can be estimated as $A/\gamma^2 \sim 0.5\times 10^{-5}$ ($\mu\Omega$cm/K$^2$)/(mJ/(moleK$^2$))$^2$ for Co$_2$MnGa, which is close to the universal value $A/\gamma^2 \sim 1\times 10^{-5}$ ($\mu\Omega$cm/K$^2$)/(mJ/(moleK$^2$))$^2$ known for strongly correlated electron systems\cite{Kadowaki1986,Tsujii2005}.

The solid curve in the main panel of Fig. S3 represents the sum of $\gamma T$ and the specific heat due to Debye phonons, namely.
\begin{eqnarray*}
C&=&\gamma T + C_{\rm D},\\
C_{\rm D}& =&9Nk_{\rm B}\frac{T^3}{T_{\rm D}^3}\int_0^{T_{\rm D}/T} \frac{x^4e^x}{(e^x-1)^2}dx,
\end{eqnarray*}
where $\gamma = 12.2$ mJ/(moleK$^2$) and $T_{\rm D}=350$ K, which is only slightly smaller than the one obtained from the fitting below 20 K. This fitting well reproduces the specific heat of Co$_2$MnGa in the wide temperature region, and confirms our low temperature estimates of both electronic and phonon contributions to the specific heat.

\section*{Density Functional Theory (DFT) calculation of electronic structure}
The electronic structure of Co$_2$MnGa was obtained by using the QUANTUM ESPRESSO package\cite{qe}, where the exchange-correlation functional within the generalized-gradient approximation by Perdew-Burke-Ernzerhof\cite{pbe} and fully-relativistic ultrasoft pseudopotentials were employed. The cutoff energy for the plane wave basis and charge density was set to 100  Ry and 1000 Ry, respectively, and a $k$-point grid of $18\times18\times18$ was used.

Following our experimental results given in the first section of Supplementary Information, the lattice constant of Co$_2$MnGa was set to 5.77 Å.
At the end of self-consistent iterations, the magnetic moment of 4.20 $\mu_{\rm B}$/f.u. was obtained, which is close to the experimentally obtained value ($\sim 4.0$~$\mu_{\rm B}$) and consistent with the previous calculations based on a full potential method and the expected value from Slater-Pauling rule\cite{Galanakis2002,guo2013}.

\section*{Fermi surfaces}
The Fermi surface of each contributing band is drawn in Fig. S4 in ascending order of energy from left to right.
The first three Fermi surfaces are closed around $\Gamma$ point, which correspond to the hole-like bands near $\Gamma$ point in Fig. 1b of the main text. 
The fourth Fermi surface, which is located near the Brillouin zone boundary and the closest to the $E_{\rm F}$, is the one which we focus on in Fig. 1 (red-colored curves in Figs. 1b-d).
The Weyl points shown in Figs. 1c and 1d are formed by the same band as the one constructing this Fermi surface.
As discussed in the main text, we believe the proximity to the Weyl points of this Fermi surface is one of the important factors to enhance the anomalous Hall conductivity at $E_{\rm F}$ (Fig. 3e in the main text).

\section*{Density of states of Co$_2$MnGa}  
The density of states near the Fermi energy is shown in Fig. S5.
The broad and sharp peaks observed at  $\sim E_{\rm F}$ and $\sim 6$ meV above $E_{\rm F}$, respectively, can contribute to enhance $\alpha_{yx}$ (main text).

\section*{Construction of Wannier representation}
From the Bloch states obtained in the DFT calculation described above, a Wannier (realistic tight-binding) basis set was constructed using the Wannier90 code\cite{w90}. The basis was composed of ($s,\  p,\  d$)- character orbitals localized at each Co and Mn site and ($s,\ p$)- character ones at Ga site, i.e., 62 orbitals/f.u. in total, including spin multiplicity.  This set was extracted after iteration for disentangling 116 bands from the space spanned by the original Bloch states in the energy range from $-11$ eV up to $+50$ eV, while freezing the bands up to $+2$ eV so as to maintain the original dispersion close to the Fermi energy. This iteration was repeated up to the point where the localization of Wannier orbitals improved no more than 0.002\% in consecutive steps. No unitary transformation for further localization was performed within the 62 orbitals. Finally, the Wannier-representation of the Hamiltonian was obtained.

\section*{Expressions of the anomalous Hall and transverse thermoelectric conductivities}
The anomalous Hall conductivity $\sigma_{yx}$ and the anomalous transverse thermoelectric conductivity $\alpha_{yx}$ that intrinsically appear in crystalline systems can be expressed with the out-of-plane component $\Omega_{n, z}({\bf k})$ of the Berry curvature as follows\cite{xiaoberry-phase2006}
\begin{align}
\sigma_{yx}(T, \ \mu) &= -\frac{e^2}{\hbar}\int\frac{d{\bf k}}{(2\pi)^3}\Omega_{n, z}({\bf k})f_{n{\bf k}}, \label{sigyx} \\
\alpha_{yx}(T, \ \mu) &= -\frac{e}{T\hbar}\int\frac{d{\bf k}}{(2\pi)^3}\Omega_{n, z}({\bf k})\{(\varepsilon_{n{\bf k}}-\mu)f_{n{\bf k}}+k_{\rm B}T\log[1+e^{-\beta(\varepsilon_{n{\bf k}}-\mu)}]\}, \label{alpyx}
\end{align}
where $e$, $\hbar$, $\varepsilon_{n{\bf k}},\ f_{n{\bf k}}$, $\mu$ are the elementary charge with negative sign, the reduced Planck constant, the band energy, and the Fermi-Dirac distribution function with the band index $n$ and the wave vector {\bf k}, and $\mu$ is the chemical potential.  
Manipulation of Eq. (\ref{alpyx}) leads to the following simple relation\cite{xiaoberry-phase2006}
\begin{equation}
\alpha_{yx}(T,\  \mu)= \frac{1}{e}\int d\varepsilon\left(-\frac{\partial f}{\partial \varepsilon}\right)\sigma_{yx}(0,\  \varepsilon)\frac{\varepsilon-\mu}{T}, \label{alpha_sigma}
\end{equation}
which can be approximated as 
 \begin{equation}
 \alpha_{yx}(T,\ \mu \simeq \varepsilon)= \frac{\pi^2k_{\rm B}^2T}{3e}\frac{\partial \sigma_{yx}(0,\ \varepsilon)}{\partial \varepsilon} \label{alpyx_lowT}
 \end{equation}
at low temperatures where thermal energy broadening is small enough compared to the Fermi energy.

The $\sigma_{yx}$ and $\alpha_{yx}$ shown in  Figs. 3b-d of the main text were computed according to Eqs. (\ref{sigyx}-\ref{alpyx}) in the Wannier representation, using, for the integrations, a $k$-point mesh of $100\times100\times100$ and additionally an adaptive mesh of $3\times3\times3$ in regions with large $\Omega_{n,\ z}$. The $\alpha_{yx}$ shown in Fig. 3e and f of the main text was evaluated by numerical differentiation corresponding to Eq.(\ref{alpyx_lowT}).

\section*{Weyl point search in the Brillouin zone}        
The Weyl points, each of them being characterized by a topological charge (Berry flux) of either $\pm 2\pi$ through a closed surface enclosing it, have been searched with our implementation of the method of Fukui-Hatsugai-Suzuki\cite{fukui2005}. 
The search was performed for each one of $50\times50\times50$ parallelepiped boxes obtained by evenly dividing the Brillouin zone: Evaluating the total Berry flux through the surface of a box, via the computation of overlaps between the Bloch states at neighboring $k$-points on a $10\times 10$ mesh on each side, it was determined whether the box contains a Weyl point, and if so, which charge it has.

\section*{Transverse thermoelectric conductivity $|\alpha_{yx}|$ of various magnets}
The values of $|\alpha_{yx}|$ plotted in Fig. 4c in the main text are taken at the temperatures where $|\alpha_{yx}|$ for each magnetic material shows the maximum in its $T$ dependence. Namely, $|\alpha_{yx}|\sim 4$ AK$^{-1}$m$^{-1}$ for Co$_2$MnGa, $|\alpha_{yx}|\sim 1$ AK$^{-1}$m$^{-1}$ for MnSi\cite{Hirokane2016}, $|\alpha_{yx}|\sim 0.6$ AK$^{-1}$m$^{-1}$ for SrRuO$_3$\cite{Miyasato2007}, $|\alpha_{yx}|\sim 0.3$ AK$^{-1}$m$^{-1}$ for Mn$_3$Sn\cite{IkhlasTomita2017,Kamran2017},
$|\alpha_{yx}|\sim 0.2,$ and $0.12$ AK$^{-1}$m$^{-1}$ for La$_{1-x}$Sr$_x$CoO$_3$ ($x=$ 0.30 and 0.25)\cite{Miyasato2007},
$|\alpha_{yx}|\sim 0.03$ AK$^{-1}$m$^{-1}$ for Nd$_2$Mo$_2$O$_7$\cite{Hanasaki2008},
$|\alpha_{yx}|\sim 0.25$ and 0.07 AK$^{-1}$m$^{-1}$ for Ga$_{1-x}$Mn$_{x}$As ($x=0.07$ and 0.05)\cite{Pu2008} and
$|\alpha_{yx}|\sim 0.0045$ AK$^{-1}$m$^{-1}$ for Fe$_3$O$_4$\cite{Ramos2014},
from the experimental reports and
$|\alpha_{yx}|\sim 0.9, 0.52$ and $0.21$ AK$^{-1}$m$^{-1}$ for FePt, Co and Fe from the calculations\cite{Weischenberg2013}. Normally, the anomalous Nernst effect is seen only in ferromagnets. Note that Fig. 4c includes one exception, the chiral antiferromagnet Mn$_3$Sn with tiny spontaneous moment $\sim3{\rm m}\mu_{\rm B}$/Mn. Interestingly, $\alpha_{yx}$ of Mn$_3$Sn is comparable to the other ferromagnets\cite{IkhlasTomita2017,Kamran2017}, because the large intrinsic contribution from the Weyl points near $E_{\rm F}$ \cite{Kuroda2017} dominates the ANE, similarly to the case of Co$_2$MnGa.

\section*{Comment on the positive magnetoconductance}
Apart from the chiral anomaly and anisotropic magnetoconductance discussed in the main text, positive longitudinal magnetoconductance (LMC) or negative longitudinal magnetoresistance (LMR) may also arise from other origins such as current jetting effect and weak localization.

In high-mobility compensated semimetals, the transverse magnetoresistance (TMR) can be very large compared to LMR, which is characterized by the large value of the ratio, $A$=TMR/LMR. For such materials, the current flows directly between the current contacts without spreading along the perpendicular direction to the field for $\vec{B} \parallel \vec{I}$, hence which is called current jetting\cite{Pippard1989}. Recently, it has been shown that this current jetting effect is the major origin for the negative LMR observed in several Weyl semimetals\cite{DosReis2016,Yuan2016b}. In Co$_2$MnGa, however, the carrier mobility is small compared to the weakly correlated semimetals and zero-gap semiconductors. Moreover, the magnetoresistance is very isotropic $A\sim 1.01$. Nevertheless, we have performed the explicit measurements to rule out the current-jetting effect.
Namely, we confirmed that the magnetoconductivity measured at one side (side-A) is almost the same as the one measured at the opposite side (side-B) over the entire range of the magnetic field up to 16 T. We found that this is indeed the case for all the current directions along [100], [110], and [111], as shown in Fig. S6a, S6b, and S6c.  Thus, we may conclude that the current jetting effect is negligible in Co$_2$MnGa.

Moreover, qualitatively the same behaviors, namely the positive magnetoconductance for $\vec{I} \parallel \vec{B}$ and the negative magnetoconductance in high magnetic field for $\vec{B} \perp \vec{I}$, are observed in the measurements with the different current directions such as $\vec{I} \parallel$ [100], $\vec{I} \parallel$ [110] and $\vec{I} \parallel$ [111] as shown in Fig. S6a, S6b, and S6c. For all the current directions, the positive LMC and negative transverse magnetoconductance becomes clear by subtracting the sharp change in magnetoconductance at low fields $B<\sim$0.3 T coming from the magnetic domain reconfiguration (Fig. S6e and S6f). 
To check the dependence on the angle between $\vec{B}$ and $\vec{I}$, we have carried out the measurements at various angles for the case of $\vec{I} \parallel$ [111] and found that the positive magnetoconductance appears only around $\vec{B} \parallel \vec{I}$, namely $\theta < \sim 10^{\circ}$) as shown in Fig. S6d. (In the case of $\theta \ge 75^{\circ}$, a higher field than 9 T is necessary to suppress spin fluctuations (see below) and to induce the negative magnetoconductance, as shown in Fig. S6c).
This can be further confirmed by the angle dependence of the magnetoconductance taken at $B=$ 9 T (Fig. 4b in the main text). The magnetoconductance exhibits $\cos^2\theta$ dependence and is indeed positive only when $\theta$ is less than $30^{\circ}$ for all the cases with $\vec{I} \parallel$ [100], [110] and [111]. These observations provide strong evidence for the chiral anomaly.

To isolate the effect due to the magnetic spin fluctuations, we have carried out the transverse magnetoconductance measurements at various temperatures with $\vec{B} \perp \vec{I} \parallel [100]$. At high temperatures, spin fluctuations are thermally induced and thus dominate the scattering process, while magnetic field may suppress such magnetic scattering. In fact, the positive transverse magnetoconductance is observed at 200 K. However, when we cooled down the system to 100 K, the positive magnetoconductance saturates around 15 T, indicating the crossover to the negative magnetoconductance at $B >$ 16 T. With further decreasing temperature down to 5 K, we observe such a crossover appears at 6 T. This confirms that the scattering due to spin fluctuations significantly weakens at low temperatures, and the positive magnetoconductance seen above 6 T for the case of $\vec{B} \parallel \vec{I}$ should not come from spin fluctuations, but from the chiral anomaly.

Field induced suppression of weak localization in conventional dirty semimetals and semiconductors is known to cause the negative magnetoresistance (positive magnetoconductance) in all directions. However, in the absence of magnetic field, weak localization causes the resistivity minimum at low temperatures\cite{Hikami1980,Kawabata1980}. We do not find such minimum in the temperature dependence of the resistivity for Co$_2$MnGa (Fig. S2). Therefore, field-induced suppression of weak localization cannot explain the observed magnetoconductance.

\section*{Anisotropic magnetoresistance and the proximity to the half metallicity}
The half metallicity and its proximity may contribute to the anisotropy in the magnetoresistance and induce so-called anisotropic magnetoresistance (AMR). While Co$_2$MnGa itself is not a half metal, the proximity is clear from its high spin polarization value $P = 0.6$ revealed by experiment\cite{Hono2009}. The AMR for Heusler alloys is extensively studied in thin films and well captured by the recent theory by Kokado $et$ $al$. based on the $s$-$d$ coupling model\cite{Sakuraba2014,Kokado2012}. The theory predicts that the systematic relation between AMR ratio, $(\rho_{I\parallel B}-\rho_{I\perp B})/\rho_{I \perp B}$, and valence electron number, $N_{\rm V}$. By using this relation, a positive AMR ratio, namely $\sigma_{I\parallel B}<\sigma_{I\perp B}$, is expected given $N_{\rm V}\sim 28$ for Co$_2$MnGa. However, we observe $\sigma_{I\parallel B}>\sigma_{I\perp B}$ (Fig. S6h), leading to the negative sign of AMR ratio, which contradicts the AMR theory and is consistent with the chiral anomaly.

\section*{Effective low energy theory}
The expression for anomalous transverse thermoelectric conductivity in Eq. (\ref{alpyx}) can also be written as 
\begin{equation}
\alpha_{xy}=\frac{k_B e}{\hbar} \; \sum_{n, k} \; \Omega^z_{n,k} \; s_{n,k},
\end{equation}
where $s_{n,k}=s(\beta[\epsilon_{nk}-\mu])=-f_{n,k}\ln(f_{n,k})-(1-f_{n,k})\ln(1-f_{n,k})$ is the entropy density for the $n$ th band, and $\beta=1/(k_B T)$. After introducing an auxiliary variable of integration $\epsilon$, we can rewrite this as 
\begin{eqnarray}
\alpha_{xy}&=&\frac{k_B e}{\hbar} \; \sum_{n, k} \; \int d\epsilon \; \Omega^z_{n,k} \delta (\epsilon-\epsilon_{nk}) \; s(\beta[\epsilon-\mu]),\\
&=& \frac{k_B}{e} \; \int d\epsilon \; \frac{\partial \sigma_{xy}}{\partial \epsilon}(\epsilon) \; s\left(\frac{\epsilon}{k_B T}\right), \label{eq2}
\end{eqnarray}
where the energy derivative of anomalous Hall conductivity $\sigma_{xy}$ is evaluated at zero temperature. 

We are interested in $\alpha_{xy}$ of a tilted, time reversal symmetry breaking Weyl semimetal, for which the right ($+$) and the left handed ($-$) Weyl fermions are described by the low energy Hamiltonians,
\begin{equation}
H_{\pm} \approx E_0 \pm \hbar v_2 (k_z \mp k_0) + \hbar v_\perp (k_x \sigma_x + k_y \sigma_y) \pm \hbar v_1 (k_z \mp k_0) \sigma_z.
\end{equation}
In the above equation $ v_1$, $v_2$ and $v_\perp$ are three independent velocity parameters, and the tilt parameter $v_2/v_1$ captures the strength of particle-hole anisotropy. For $v_2/v_1 <1$, we obtain a type I Weyl semimetal with only one of the energy bands (conduction or valence) producing Fermi pockets, while the density of states vanishes at the touching point (located at a reference energy $E_0$). On the other hand, for $v_2/v_1>1$ we find a type II Weyl metal, where both energy bands can produce Fermi pockets, and the electron and hole like pockets touch at the Weyl points. Consequently, the type II Weyl fermions possess a finite density of states at the touching point. These two types of Weyl fermions are separated by the critical tilt parameter $v_2/v_1=1$, which corresponds to a quantum Lifshitz critical point. 

If the chemical potential remains tuned at the touching points (i.e, $\mu=E_0$), the quantum Lifshitz point describes a phase transition between an incompressible (zero density of states) Weyl semimetal and a compressible Weyl metal. However, the topological properties for any $v_2/v_1$ remain unaffected, as the coefficients of the three Pauli matrices are kept unchanged. At this critical point, the energy derivative of the anomalous Hall conductivity $\frac{\partial \sigma_{xy}}{\partial \epsilon}(\epsilon)$ shows divergent or singular scaling behavior at low energies. While approaching the critical point, $\xi \sim 1/(k_0 |\delta|)$ denotes a divergent correlation length, where $\delta=1-v_2/v_1$ is the reduced distance from the critical point. Therefore, in the vicinity of the Lifshitz point we can write the following scaling form 
\begin{equation}
\frac{\partial \sigma_{xy}}{\partial \epsilon}(\epsilon) =\frac{e^2}{4 \pi^2 \hbar^2 v_1}F \left( \frac{\epsilon}{|\delta| E_c} \right)
\end{equation}
where, $F(x)$ is a dimensionless scaling function, and $E_c=\hbar v_1 k_0$ is the high-energy cutoff for the low energy theory of Weyl fermions. Consequently, the temperature and chemical potential dependence of $\alpha_{xy}$ can be captured in terms of a scaling function 
\begin{equation}
\alpha_{xy}(k_BT,\mu, \delta) = \frac{k^2_B e T}{4 \pi^2 \hbar^2 v_1} G \left( \frac{k_B T}{|\delta| E_c},\frac{\mu}{|\delta| E_c} \right).
\end{equation}
Through detailed analytical and numerical calculations, we can show that at low energies the energy derivative of the Hall conductivity shows a logarithmic singularity i.e., $\frac{\partial \sigma_{xy}}{\partial \epsilon} \sim \log \left(\frac{|\epsilon|}{\hbar v_1 k_0}\right)$ when $\epsilon \ll \hbar v_1 k_0$. When the system is tuned away from the critical point, this singularity would be rounded off by the finite correlation length, leading to $ \frac{\partial \sigma_{xy}}{\partial \epsilon} \sim \log(|\delta|)$ behavior for $\epsilon \ll \delta \hbar v_1 k_0$. Nevertheless in the vicinity of the critical point, the logarithmic singularity manifests itself over a wide range of energies or the quantum critical fan defined as $\delta \hbar v_1 k_0 < \epsilon < \hbar v_1 k_0$. Notably, the logarithmic divergence of $\frac{\partial \sigma_{xy}}{\partial \epsilon}$ at the critical point would lead to $\alpha_{xy} \sim T \log (|E_F-E_0| /(\hbar v_1 k_0)) $ behavior at low temperatures. But at high temperatures $k_B T> |E_F-E_0|$, the logarithmic singularity would give rise to $\alpha_{xy} \sim T \log(k_B T/(\hbar v_1 k_0))$ behavior, which captures the quantum critical violation of Mott formula ( $\alpha_{xy} \sim T$). We emphasize that the Mott formula will be violated at arbitrarily low temperatures, if the chemical potential is tuned precisely at the touching point ($\mu=E_0$).

The first-principles calculations for Co$_2$MnGa show the existence of a pair of type-I Weyl fermions with $k_0 \sim 0.15  \times 2\pi/a$ located around $E_0 \approx +0.02$ eV, with a tilt parameter $v_2/v_1=0.99$ (Fig. 1 in the main text). These Weyl points are on the verge of undergoing a quantum Lifshitz transition, with $\delta=0.01$. Consequently, the characteristic property of a type-I Weyl fermion can only be realized at very low energies when $|E_F-E_0|$ or $k_B T$ are much smaller than $\delta \hbar v_1 k_0 \sim$ 1 meV. Since the distance of the Weyl points from the Fermi level $|E_F-E_0| \sim 20$ meV is larger than $\delta \hbar v_1 k_0$, but smaller than $\hbar v_1 k_0 \sim 0.1$ eV, we are actually accessing the quantum critical fan of the Lifshitz point. Therefore, we need to calculate the crossover between $k_BT< \mu$ and $k_B T>\mu$ regimes by essentially restricting ourselves to the quantum critical point. 
 
At the Lifshitz point, the band dispersions along the nodal separation become flat. In particular, one of the bands can produce a very large Fermi pocket along the nodal direction, and it is important to employ a lattice regularization along the nodal separation for avoiding any spurious divergence at large energies or momenta. Hence, we would consider the following approximate two-band model
\begin{equation}
H \approx (E_0+t_2[\cos(k_z a/2)-\cos(k_0 a/2)]) t+v_\perp[k_x \sigma_x + k_y \sigma_y] + t_1[\cos(k_z a/2)-\cos(k_0 a/2)]\sigma_z,
\end{equation}
with $t_1=t_2=t$, leading to $v_1=v_2=v=t a/(2\hbar)\sin(k_0 a/2)$. For this lattice regularized theory, the scaling functions can be suitably defined as
\begin{eqnarray}
&&\frac{\partial \sigma_{xy}}{\partial \epsilon} = \frac{e^2}{2\pi^2 \hbar ta} \; F\left(\frac{\epsilon}{t}, k_0a \right), \: \: \alpha_{xy}(k_BT/t,\mu/t)=\frac{k^2_B eT}{2\pi^2 \hbar ta} G\left(\frac{k_BT}{t}, \frac{\mu}{t} \right), \\
&&G\left(\frac{k_BT}{t}, \frac{\mu}{t} \right)=\frac{t}{k_BT} \; \int dx F(x, k_0a) \; s \left(\frac{t[x-\mu/t]}{k_BT} \right).
\end{eqnarray}
Now we proceed with the evaluation of $F(x,k_0a)$. For $x>0$, only the conduction band with energy dispersion $\epsilon_{+,k}=t[\cos(k_z a/2)-\cos(k_0a/2)] + \sqrt{\hbar^2v^2_\perp k^2_\perp+ t^2 [\cos(k_za/2)-\cos(k_0 a/2)]^2}$ can produce a Fermi pocket. In the low energy regime $x=\epsilon/t< 2[1-\cos(k_0a/2)]$, the $k_z$ coordinate for the Fermi pocket will be bounded by $\arccos[\cos (k_0a/2) + x/2] < |k_z a/2| < \pi$. Therefore, no portion of electron-like pocket would be located in the region $-\arccos[\cos (k_0a/2) + x/2]<k_z a/2 <\arccos[\cos (k_0a) + x/2]$. Only in this energy regime, the low energy theory of Weyl fermions can be meaningful. By contrast, for sufficiently large energies satisfying $x> 2[1-\cos(k_0a/2)]$, the electron like pocket would cover the entire length of the Brillouin zone $0<|k_z a|< 2 \pi$. Similarly, we can show that for $x<0$ but $x>-2[1+\cos(k_0a/2)]$, we only obtain a hole pocket in the region $-\arccos[\cos (k_0a/2) + x/2]< k_z a/2 < \arccos[\cos (k_0a) + x/2]$. All the integrals for determining $F$ can be performed analytically. However, the actual forms of $F$ are not very illuminating. If we concentrate on $|x|< [1-\cos(k_0a/2)]$, 
\begin{eqnarray}
F(x>0,k_0a)=\int^{\pi}_{z_0} \; dz \; \frac{\cos(z)-\cos(k_0a/2)}{[x+\cos(k_0a/2)-\cos(z)]^2}, \nonumber \\
F(x<0,k_0a)=-\int^{z_0}_{0} \; dz \; \frac{\cos(z)-\cos(k_0a/2)}{[x+\cos(k_0a/2)-\cos(z)]^2}, \nonumber \\
\end{eqnarray}
with $z_0=\arccos[\cos (k_0a/2) + x/2]$. After finishing the integral over $z$, we find that for small $x$,
$$F \sim \frac{1}{\sin(k_0a/2)} \log \left(\frac{|\epsilon-E_0|}{C(k_0) \hbar v_1 k_0} \right)$$ as announced in the Eq. (1) of the main text. This logarithmic dependence is clearly identified in the first-principles calculation. After determining $F$, we can perform the integration over $x$ numerically to obtain $\alpha_{xy}$, which shows the crossover between the $T$-linear (at low temperatures $\mu> k_BT$) and $T \log(T/T_0)$ (at high temperatures $k_BT_0> k_B T> \mu$) behaviors. However, the integrations for $E_F=E_0$, can be done analytically to obtain the results in Eq. (3).  By plotting the energy derivative of the anomalous Hall conductivity determined by the first-principles calculation around $E=+0.02$ eV against $\log(E-E_0)$ in Fig. 3f, we establish the $\log(|\epsilon|)$ scaling behavior of the underlying Weyl fermions. Similar logarithmic behaviors are also found for the other extrema of $\frac{\partial \sigma_{xy}}{\partial \epsilon}$. This indicates the existence of additional pairs of Weyl points which also belong to the quantum critical regime. The contributions of all such pairs can enhance the predictions of effective theory by a factor of $N_f$, where $N_f$ is the number of such pairs. The experimentally and numerically found $\alpha^{m}_{xy}$ can be obtained with $N_f \sim 3$. We have also verified that the calculations for the full lattice model only changes the estimation for $T_0$, but does not cause any significant variation of $\alpha^{m}_{xy}$.

\section*{2. Figure of Merit}
For thermoelectric materials based on the Seebeck effect, the dimensionless figure of merit $ZT=\sigma_{xx}S_{xx}^2T/\kappa_{xx}$ is usually used to estimate the efficiency because the maximum efficiency is described by $ZT$,
\begin{eqnarray*}
\eta_{\rm SE}&=&\eta_{\rm C} \frac{\sqrt{1+ZT_{\rm av}}-1}{\sqrt{1+ZT_{\rm av}}+T_{\rm c}/T_{\rm h}},\\
\end{eqnarray*}
where $T_{\rm c}$ and $T_{\rm h}$ are temperatures at the hot and cold parts, $T_{\rm av}=(T_{\rm h}+T_{\rm c})/2$ is the average temperature
and $\eta_{\rm C}=(T_{\rm h}-T_{\rm c})/T_{\rm h}$ is Carnot efficiency, respectively \cite{Goldsmid2009}.

On the other hand, the efficiency for the thermoelectric generation by Nernst effect $\eta_{\rm ANE}$ can be written as 
\begin{eqnarray*}
\eta_{\rm ANE}&=&\eta_{\rm C} \frac{1-\sqrt{1-Z_{yx}T_{\rm av}}}{1+\sqrt{1-Z_{yx}T_{\rm av}}},\\
\end{eqnarray*}
where $Z_{yx}T=\sigma_{xx}S_{yx}^2T/\kappa_{yy}$ is the figure of merit for ANE \cite{Harman1962}. Since the functional form of $\eta_{\rm ANE}$ is very different from that of $\eta_{\rm SE}$, e.g. $\eta_{\rm ANE}\rightarrow \eta_{\rm C}$ at $Z_{yx}T\rightarrow 1$ while $\eta_{\rm ANE}\rightarrow \eta_{\rm C}$ at $ZT\rightarrow \infty$, $Z_{yx}T$ cannot be directly compared to $ZT$. 

For Co$_2$MnGa, we have estimated $Z_{yx}T$ as shown in Fig. S7. $Z_{yx}T$ is nearly isotropic, which is beneficial for applications. $Z_{yx}T$ monotonically increases on heating and reaches $Z_{yx}T\sim 0.08$ ($\%$) at $T= 400$ K. This is much smaller compared to the reported values for Seebeck effect and comparable to those for spin Seebeck effect \cite{Uchida2016}.


\newpage

\begin{figure}[t]
\begin{center}
\includegraphics[keepaspectratio, scale=0.45]{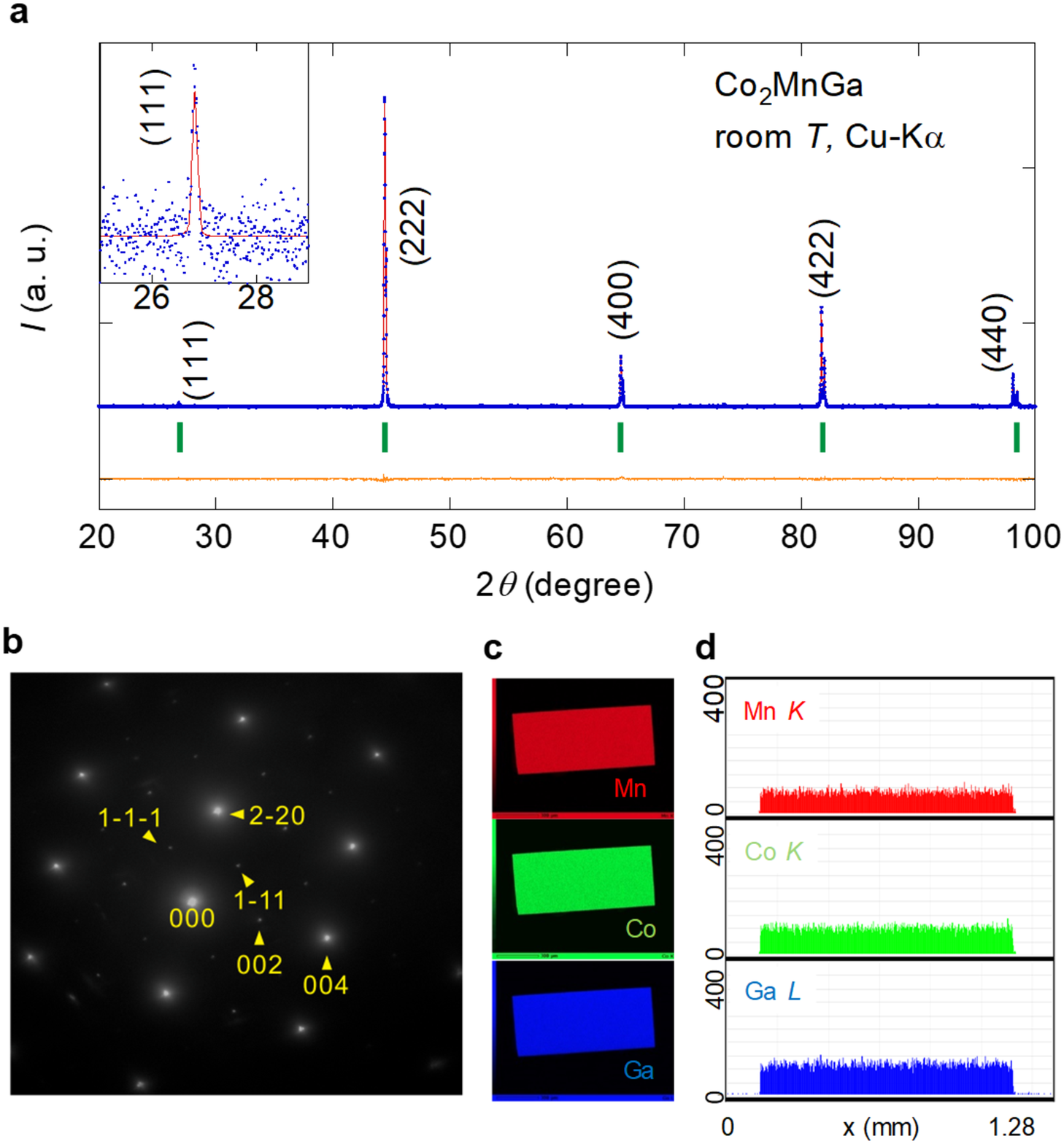}
\end{center}
\end{figure}
\normalsize{
{\bf Figure S1 $|$ Evidence for $L2_1$ structure and the homogeneity of the crystal.} {\bf a} X-ray diffraction (XRD) pattern of Co$_2$MnGa obtained by Cu K$\alpha$ ($\lambda = 1.5401$ \AA) radiation at room temperature.  The circles and the solid line (red) represent the experimental results and the Rietveld refinement fit, respectively. The final $R$ indicators are $R_{\rm WP}=0.69$ $\%$, $R_{\rm e}=0.65$ $\%$ and $S=1.06$. Vertical bars (green) below the curves indicate the peak positions of the Co$_2$MnGa full Heusler $L2_1$ phase. The lower curve (orange) represents the difference between the experimental result and the Rietveld refinement. {\bf b} Selected area electron diffraction pattern for our single crystal of Co$_2$MnGa along [110]. The reflections of $h + k = 2n$ and $k + l = 2n$ for $hkl$ were observed as the reflections equivalent to [111]. The reflections equivalent to [200] appeared owing to multiple diffraction. {\bf c-d} 2D- ({\bf c}) and 1D- ({\bf d}) X-ray maps of Mn $K$-, Co $K$-, and Ga $L$- lines for the single crystal Co$_2$MnGa.
}

\newpage
\begin{figure}[t]
\begin{center}
\includegraphics[keepaspectratio, scale=0.7]{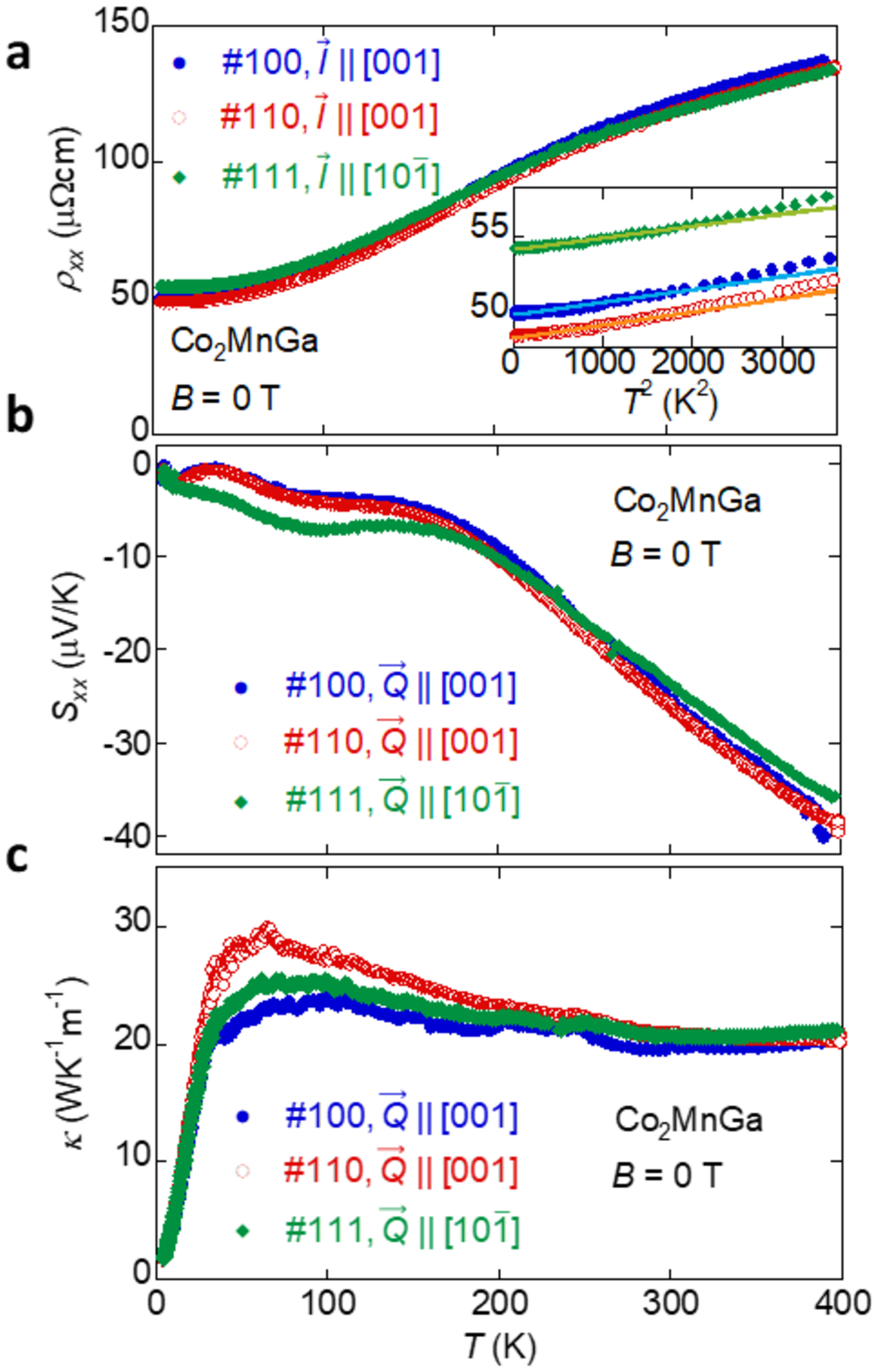}
\end{center}
\end{figure}
\normalsize{{\bf Figure S2 $|$ Electric and thermal transport properties for Co$_2$MnGa.} {\bf a-c}, Temperature dependence of the  longitudinal electric resistivity ({\bf a}), Seebeck effect ({\bf b}) and thermal conductivity $\kappa$ ({\bf c}) measured at $B$ = 0 with different heat current ($\vec{Q}$) directions.  Inset of {\bf a} shows $T^2$ dependence of the longitudinal electric resistivity $\rho_{xx}$. The solid lines indicate linear fits.
}

\newpage
\begin{figure}[t]
\begin{center}
\includegraphics[keepaspectratio, scale=0.7]{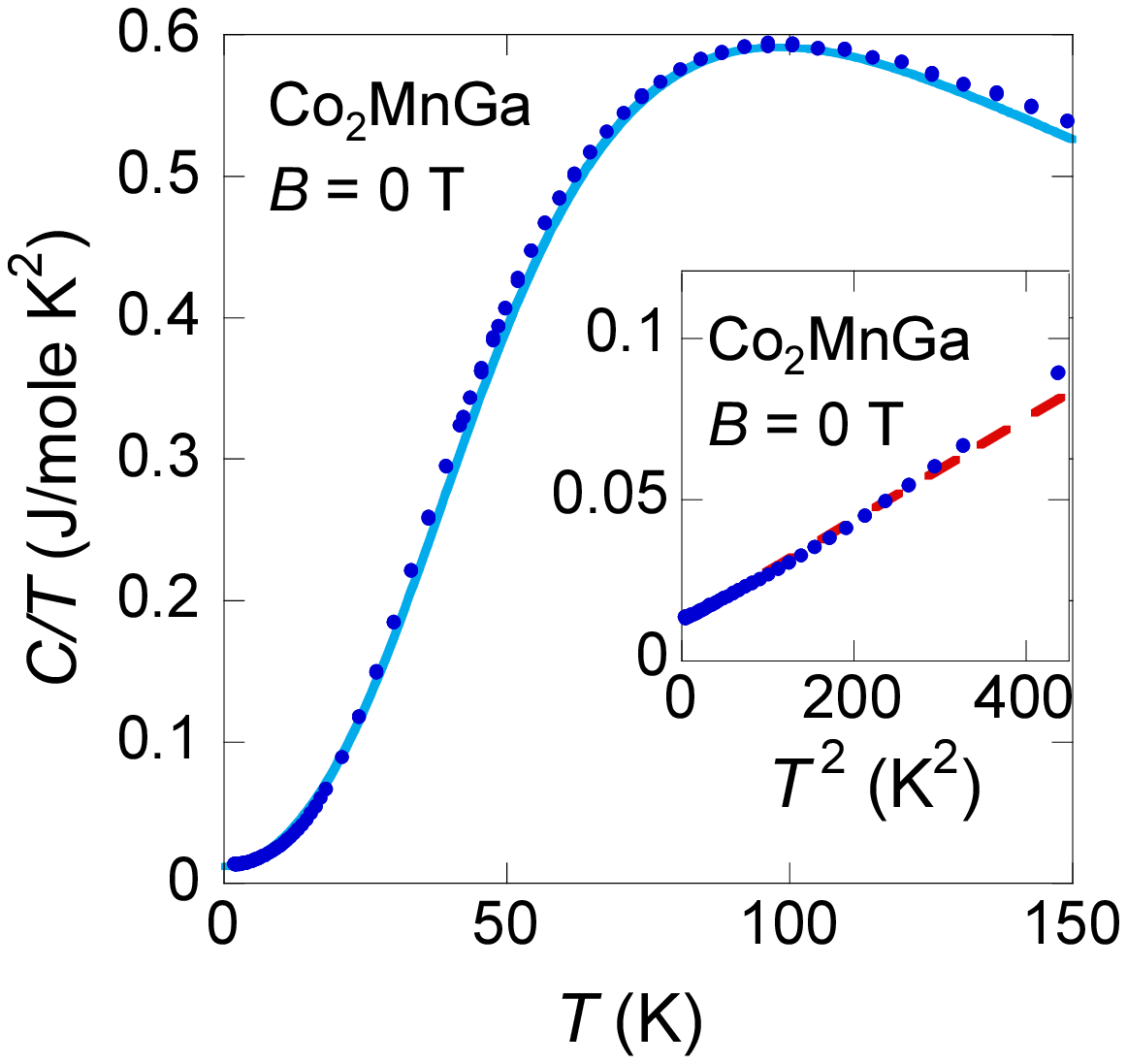}
\end{center}
\end{figure}
\normalsize{{\bf Figure S3 $|$ Specific heat for Co$_2$MnGa at low temperatures.}  Temperature dependence of the specific heat divided by $T$, $C/T$, for Co$_2$MnGa measured at $B = 0$ T. The solid curve indicates the fit to the Debye formula. The inset shows the plot of $C/T$ vs. $T^2$ below $T\sim20$ K. The broken line indicates a linear fit.}

\newpage
\begin{figure}[t]
\begin{center}
\includegraphics[keepaspectratio, scale=0.2]{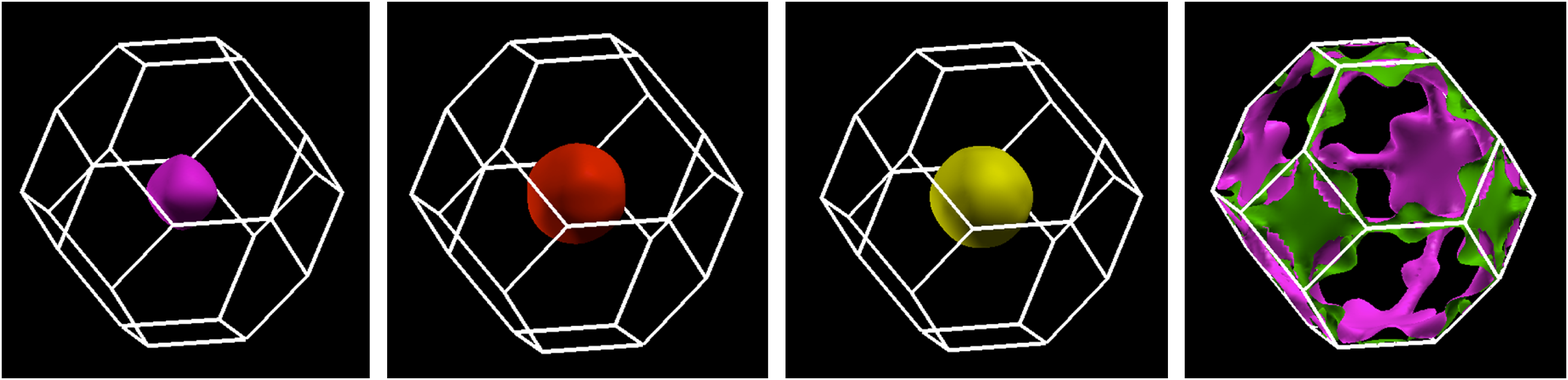}
\end{center}
\end{figure}
\normalsize{{\bf Figure S4 $|$ Fermi surfaces for Co$_2$MnGa.} Fermi surfaces of each band obtained from the first-principles calculation for the case of magnetization of 4.2$\mu_{\rm B}$ along [110] direction (in ascending order of energy).}

\newpage
\begin{figure}
\begin{center}
\includegraphics[keepaspectratio,scale=0.9]{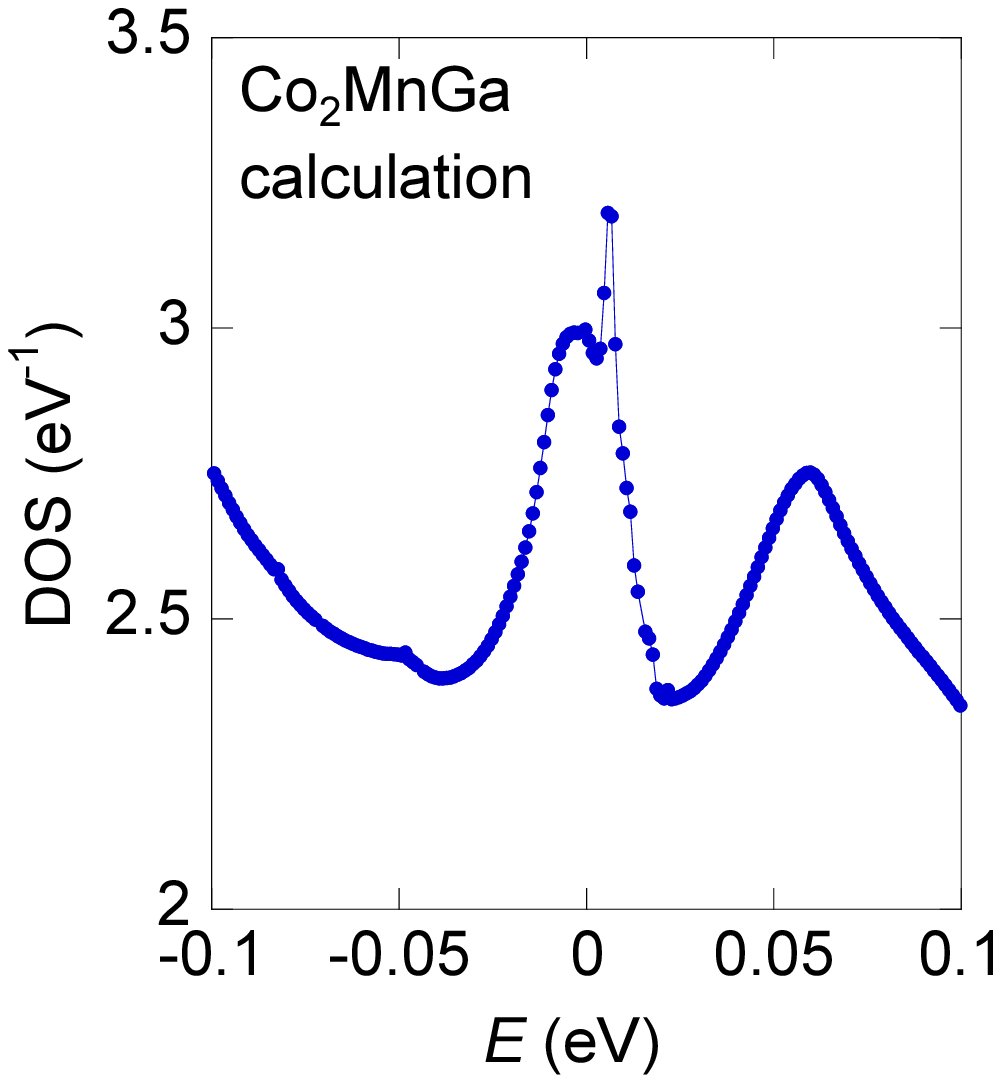}
\end{center}
\end{figure}
\normalsize{{\bf Figure S5 $|$ Density of states for Co$_2$MnGa obtained from the first-principles calculations.} Density of states around the Fermi energy $E_{\rm F}$ for the case with a magnetization of 4.2 $\mu_{\rm B}$ along [110] direction. A $k$-point grid of 50$\times$50$\times$50 and an energy smearing of 1 meV were employed. }

\newpage
\begin{figure}[t]
\begin{center}
\includegraphics[keepaspectratio, scale=0.37]{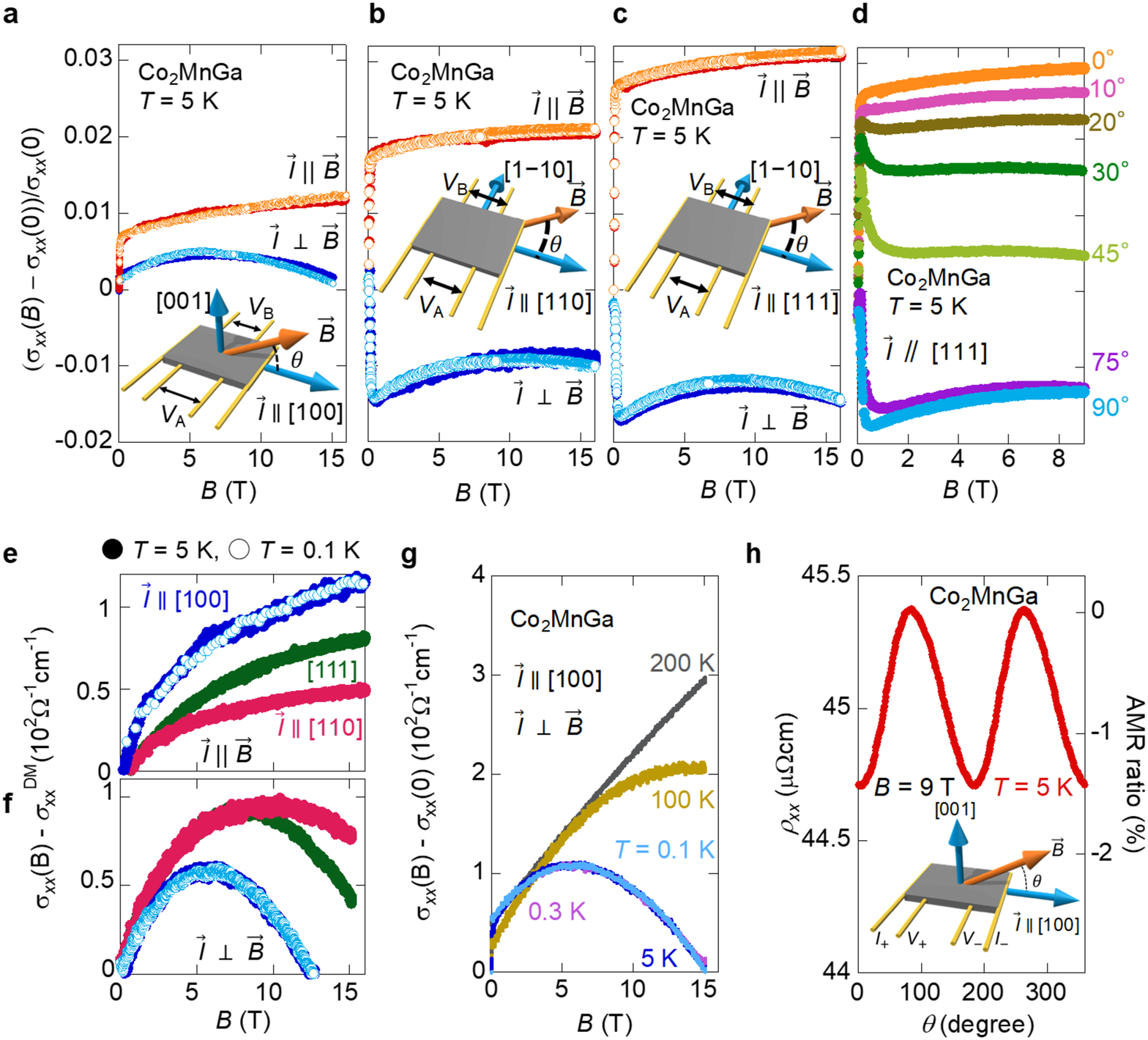}
\end{center}
\end{figure}
\normalsize{{\bf Figure S6 $|$ Absence of the current jetting effect.} {\bf a-c},  Field dependence of the magnetoconductivity $(\sigma_{xx}(B)-\sigma_{xx}(0))/\sigma_{xx}(0)$ measured using the two different voltage terminals on the side-A (solid circles) and the side-B (open circles) for $\vec{I} \parallel$ [100] ({\bf a}), $\vec{I} \parallel$ [110] ({\bf b}) and $\vec{I} \parallel$ [111] ({\bf c}). $\theta = 0$ and 90$^{\circ}$ correspond to the configurations of $\vec{I} \parallel \vec{B}$ and $\vec{I} \perp \vec{B}$, respectively. The insets show the schematic pictures of the magnetoconductivity measurement set-up to illustrate the voltage $V$ terminal positions at two different sides (side-A and side-B) of a single crystal. Here, $\theta$ is the angle between the magnetic field $\vec{B}$ and the electric current $\vec{I}$. {\bf d,} $B$ dependence of the magnetoconductivity at different angles $0^{\circ} \leq \theta \leq 90^{\circ}$.
{\bf e, f,} $B$ dependence of $\sigma_{xx}(B)-\sigma_{xx}(0)$ measured with $\vec{I} \parallel \vec{B}$ ({\bf e}) and $\vec{I} \perp \vec{B}$ ({\bf f}) and with different current directions, namely, $\vec{I} \parallel$ [100] (blue), $\vec{I} \parallel$ [110] (red) and $\vec{I} \parallel$ [111] (green).
 {\bf g}, $B$ dependence of $\sigma_{xx}(B)-\sigma_{xx}(0)$ measured at various temperatures for $\vec{I} \parallel$ [100]  $\perp \vec{B}$. {\bf h}, Anisotropic magnetoresistance (AMR) ratio of Co$_2$MnGa as a function of the angle $\theta$ at $|\vec{B}| = 9$ T. AMR ratio is defined in the text of Supplementary Information. $\theta$ is the angle between the magnetic field and the electric current direction [100], as schematically shown in the inset. }

\newpage
\begin{figure}[t]
\begin{center}
\includegraphics[keepaspectratio,scale=0.6]{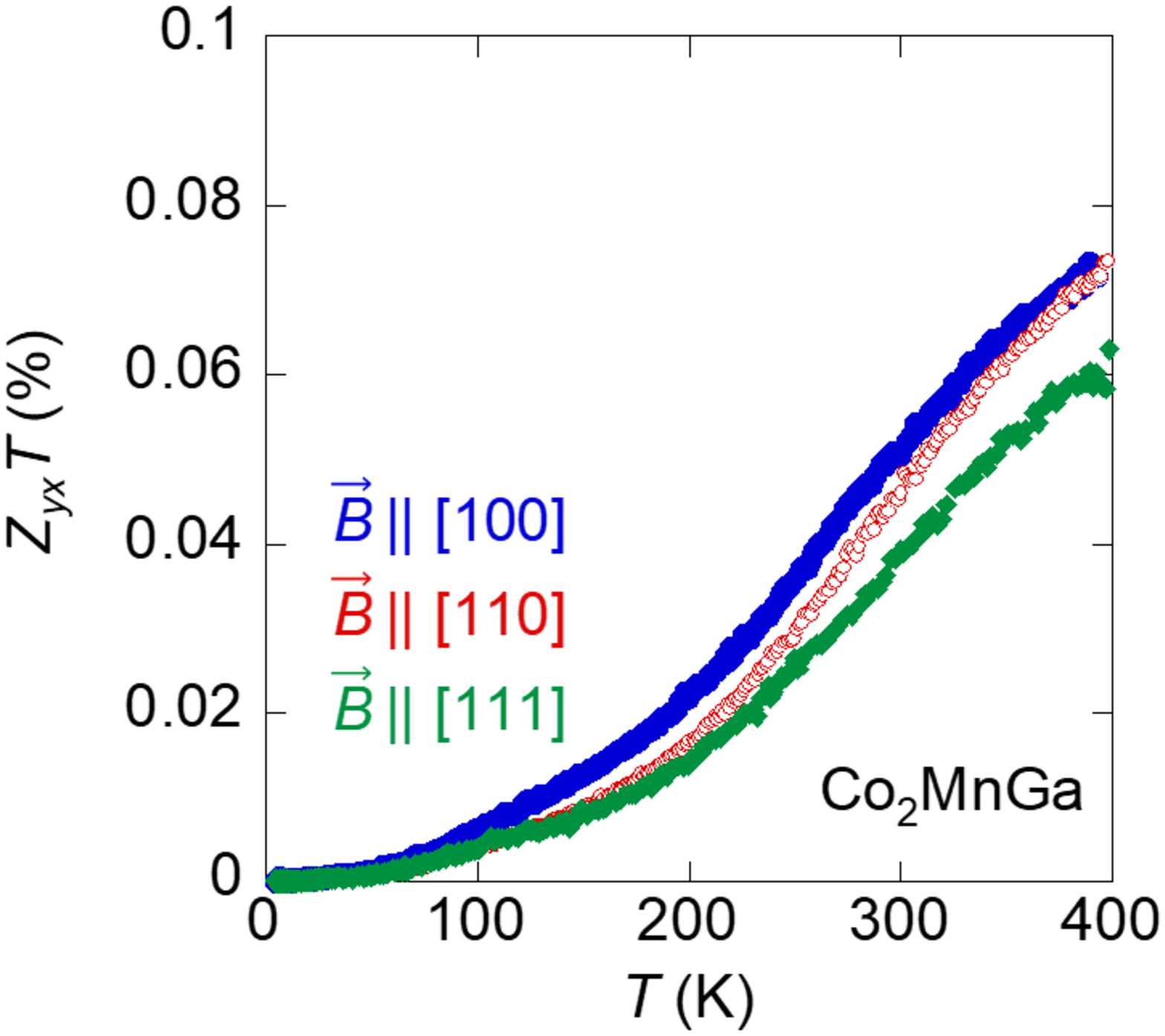}
\end{center}
\end{figure}
\normalsize{
{\bf Figure S7$|$ Figure of merit for the Nernst effect of Co$_2$MnGa.}  Temperature dependence of the figure of merit $Z_{yx}T$ for the Nernst effect measured in the magnetic field along [100] (blue), [110] (red), and [111] (green) for Co$_2$MnGa.
}

\newpage
\normalsize{
{\bf Table S1 $|$  Crystal structure parameters for Co$_2$MnGa at room temperature refined by Rietveld analysis.} The lattice parameter is determined by the analysis for the X-ray diffraction spectra with  Cu K$\alpha$ radiation.
}
\begin{table}
\begin{center}
\begin{tabular}{|l|l|l|l|l|l|}
\hline
\multicolumn{6}{|l|}{lattice parameter 5.77268 \AA}          \\ \hline
Atom & Wyckoff position & x    & y    & z    & Occupancy \\ \hline
Co   & 8c               & 0.25 & 0.25 & 0.25 & 1         \\ \hline
Mn   & 4a               & 0    & 0    & 0    & 1         \\ \hline
Ga   & 4b              & 0.5  & 0.5  & 0.5  & 1         \\ \hline
\end{tabular}
\end{center}
\end{table}

\end{document}